\documentclass[aps,prb,twocolumn,a4paper,showpacs,superscriptaddress,groupedaddress]{revtex4-1} 
\usepackage{amsmath}
\usepackage{amssymb}
\usepackage{bm}
\usepackage[usenames]{color}
\usepackage[normalem]{ulem}
\usepackage[dvips]{graphicx} 

\definecolor{myorange}{rgb}{0.7,0.5,0.0}
\definecolor{mygreen}{rgb}{0.0,0.7,0.0}
\definecolor{purple}{rgb}{0.75,0.0,1.0}

\newcommand{\tr}[1]{\textcolor{black}{#1}}
\newcommand{\trr}[1]{\textcolor{black}{#1}}
\newcommand{\tred}[1]{\textcolor{black}{#1}}
\newcommand{\tb}[1]{\textcolor{black}{#1}}
\newcommand{\tm}[1]{\textcolor{black}{#1}}
\newcommand{\tor}[1]{\textcolor{black}{#1}}

\newcommand{\tbb}[1]{\textcolor{black}{#1}}
\newcommand{\tg}[1]{\textcolor{black}{#1}}
\newcommand{\tc}[1]{\textcolor{black}{#1}}

\newcommand{\tp}[1]{\textcolor{black}{#1}}

\begin{document}
  \title{{\it Ab initio} effective Hamiltonians for cuprate superconductors}
  \author{Motoaki Hirayama$^{1)}$, Youhei Yamaji$^{2)}$, Takahiro Misawa$^{3)}$ and Masatoshi Imada$^{2)}$}
  \affiliation{$^{1)}$Department of Physics, Tokyo Institute of Technology, Japan}
  \affiliation{$^{2)}$Department of Applied Physics, University of Tokyo, 7-3-1 Hongo, Bunkyo-ku, Tokyo 113-8656, Japan}
    \affiliation{$^{3)}$Institute for Solid State Physics, University of Tokyo, Kashiwanoha, Kashiwa, Chiba, Japan}
    
\begin{abstract}
\tred{{\it Ab initio} low-energy effective Hamiltonians of two typical high-temperature copper-oxide superconductors, whose mother compounds are La$_2$CuO$_4$ and HgBa$_2$CuO$_4$, are derived by utilizing the multi-scale {\it ab initio} scheme for correlated electrons (MACE).  The effective Hamiltonians obtained in the present study serve as platforms of future studies to accurately solve the low-energy effective Hamiltonians beyond the density functional theory. It allows further study on the superconducting mechanism from the first principles and quantitative basis without adjustable parameters not only for the available cuprates but also for future design of higher $T_{\rm c}$ in general. More concretely, we derive effective Hamiltonians for three variations, 1)  one-band Hamiltonian for the antibonding orbital generated from strongly hybridized Cu $3d_{x^2-y^2}$ and O $2p_{\sigma}$ orbitals  
2) two-band Hamiltonian constructed from the antibonding orbital and Cu $3d_{3z^2-r^2}$ orbital hybridized mainly with the apex oxygen $p_z$ orbital 
3) three-band Hamiltonian consisting mainly of Cu $3d_{x^2-y^2}$ orbitals and two O $2p_{\sigma}$ orbitals.   Differences between the Hamiltonians for  La$_2$CuO$_4$ and HgBa$_2$CuO$_4$, which have relatively low and high critical temperatures $T_{\rm c}$, respectively, at optimally doped compounds, are elucidated. The main differences are summarized as 
i) the oxygen $2p_{\sigma}$ orbitals are farther ($\sim 3.7$ eV) below from the Cu $d_{x^2-y^2}$ orbital in case of the  La compound than the Hg compound ($\sim 2.4$ eV) in the three-band Hamiltonian. This causes a substantial difference in the character of the $d_{x^2-y^2}$-$2p_{\sigma}$ antibonding band at the Fermi level and makes the effective onsite Coulomb interaction $U$ larger  for the La compound than the Hg compound for the two- and one-band Hamiltonians. ii) The ratio of the second-neighbor to the nearest transfer $t'/t$ is also substantially different (\tg{0.26} for the Hg and \tg{0.15} for the La compound) in the one-band Hamiltonian. 
Heavier entanglement of the two bands in the two-band Hamiltonian implies that the 2-band rather than the 1-band Hamiltonian is more appropriate  for the La compound. 
The relevance of the three-band description is also discussed especially for the Hg compound.}
\end{abstract}

\maketitle
\section{Introduction}
Superconductors that have high $T_{\rm c}$ hopefully above room temperature at ambient pressure are a holy grail of physics. Thirty years ago, an important step forward has been made by the discovery of copper oxide superconductors\cite{Bednorz}, which have raised the record of $T_{\rm c}$ more than 100K up to around 138K\cite{Dai1995} at ambient pressure and around 160K under pressure\cite{Nunezregueiro1993,Gao1994}. However, the highest $T_{\rm c}$ record has not been broken much since then, except recent discovery of $T_{\rm c}\sim 200$K in hydrogen sulfides at extremely high pressure($>150$GPa)\cite{H3S}.  

Despite hundreds of proposals, the mechanism of superconductivity in the cuprates has long been the subject of debate and still remains as an open issue.  If the mechanism could be firmly established, the materials design for higher  $T_{\rm c}$ would greatly accelerate. In this respect, first-principles calculations of the electronic structure based on faithful experimental conditions and the quantitative reproduction of the experimental results together are a crucial first step, for the predictive power for real materials in the next step. 

From the early stage after the discovery of the cuprate superconductors, the electronic structures have been studied based on the conventional local density approximation of the density functional approach\cite{Mattheiss1987,Massidda1987,Pickett1991}.  However, the cuprate superconductors belong to typical strongly correlated electron systems\cite{Anderson1987}, which makes the conventional approach by the density functional theory (DFT) questionable. 

Theoretical studies postulating strong electron correlations have been pursued to capture the mechanism of the superconductivity more or less independently of the first principles approaches.  Those start from the Hubbard-type effective models or other simple strong coupling effective Hamiltonians with diverse and sometimes contradicting views spreading from weak coupling scenario such as spin fluctuation theory to strong coupling limit assuming the local Coulomb repulsion as the largest parameter.  Although rich concepts have emerged from diverse studies emphasizing different aspects of the electron correlation, the relevance and mechanism working in the real materials are largely open.   
This screwed up front urges the first-principles study that allows quantitative and accurate treatments of strong electron correlations without adjustable parameters.   The significance of {\it ab initio} studies is particularly true for strongly correlated systems in general, because they are subject to strong competitions among various orders and a posteriori theory with adjustable parameters does not have predictive power. There exists earlier attempts to extract parameters of effective Hamiltonians from the density functional theory~\cite{Hybertsen1989}.

To make a systematic approach possible along this line, multi-scale {\it ab initio} scheme for correlated electrons (MACE) has been pursued and developed \cite{ImadaMiyake}. MACE has succeeded in reproducing the phase diagram of the iron based superconductors basically on a quantitative level without  adjustable parameters, particularly for the emergence of the superconductivity and antiferromagnetism separated by electronic inhomogeneity\cite{misawa2012,misawa2014}. This is based on the solution of an {\it ab initio} effective Hamiltonian\cite{Miyake2010} for the five iron $3d$ orbitals derived from the combination of the density functional theory (DFT) calculations and the constrained random phase approximation (cRPA)\cite{aryasetiawan04}. 

In this paper, we apply essentially the same scheme to derive the {\it ab initio} effective Hamiltonian for two examples of the mother materials of the cuprate superconductors, La$_2$CuO$_4$ and HgBa$_2$CuO$_4$ and compare their differences. One aim of the present work is to understand distinctions of the two compounds which show contrasted maximum critical temperature at optimum hole doping (40K for La$_2$CuO$_4$ and 90K for HgBa$_2$CuO$_4$). 
The present study also serves as a platform and springboard to future studies to solve the {\it ab initio} effective Hamiltonians derived here by accurate solvers.   

In the present application of the MACE, we employ more refined scheme\cite{hirayama13,hirayama15,hirayama17} by replacing the cRPA with the constrained GW (cGW) approximation to remove the double counting of the correlation effects in the procedure of solving the effective Hamiltonian on top of the exchange correlation energy in the DFT that already incompletely takes into account the electron correlation. In the cGW scheme, effects from the exchange correlation energy contained in the initial DFT band structure is completely removed and replaced by the GW self-energy, which takes into account only the contribution from the Green's function in the Hilbert space outside of the low-energy effective Hamiltonian.  The main part of the correlation effects arising from the low-energy degrees of freedom is completely ignored at this stage and will be considered when one solves the  low-energy effective Hamiltonian beyond LDA and GW. 

Our scheme is supplemented by the self-interaction correction (SIC) to remove the double counting in the Hartree term, (or in other words, to recover the cancellation of the self-interaction between that contained in the Hartree term and that in the exchange correlation held in the LDA, but violated when only the exchange correlation is subtracted). 

\begin{figure}[ptb]
\centering 
\includegraphics[clip,width=0.4\textwidth ]{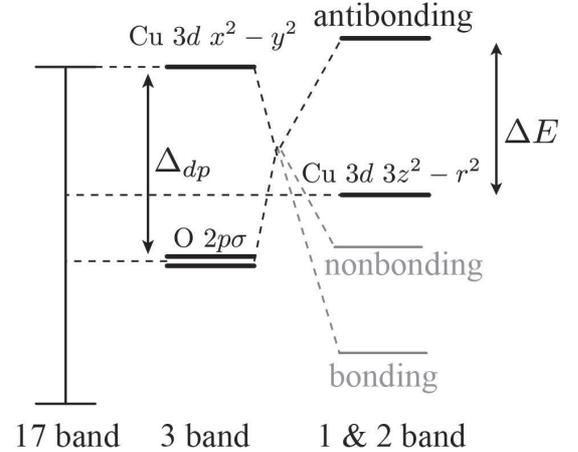} 
\caption{(Color online)
\tp{
Schematic energy \tc{levels} of orbitals \tc{constituting} three effective Hamiltonians.}
}
\label{Schematic_level}
\end{figure} 
\tp{We derive three effective Hamiltonians for La$_2$CuO$_4$ and HgBa$_2$CuO$_4$ by using the cGW scheme supplemented by SIC.
These {\it ab initio} effective Hamiltonians extract sub-Hilbert spaces expanded by combinations of Cu $3d$ $x^2-y^2$, Cu $3d$ $3z^2 -r^2$,
and O $2p_{\sigma}$ orbitals(, which is schematically illustrated in Fig.~\ref{Schematic_level})
The present downfolding scheme to derive these Hamiltonians consists of two steps: First, a 17-band effective Hamiltonian is derived.
Then, the three effective low-energy Hamiltonians are derived from the 17-band Hamiltonian hierachically.
Here, the three effective Hamiltonians are an one-band Hamiltonian for the antibonding orbital generated from hybridized Cu $3d$ $x^2-y^2$ and O $2p_{\sigma}$ orbitals,
a two-band Hamiltonian constructed from the antibonding orbital and Cu $3d$ $3z^2-r^2$ orbital hybridized mainly with the apex oxygen $p_z$ orbital,
a three-band Hamiltonian consisting mainly of Cu $3d$ $x^2-y^2$ orbitals and two O $2p_{\sigma}$ orbitals.}
\tr{\tp{A s}ummary of the obtained important \tp{matrix elements of the three effective Hamiltonians} in the present work is listed in Table~\ref{parameter_summary}.}
\tp{There are two important energy scales in the one-body part of the derived effective Hamiltonians,
in addition to the differece in effective Coulomb repulsion:
Energy difference between the oxygen $2p_{\sigma}$ orbitals and the copper $3d$ $x^2-y^2$ orbital ($\Delta_{dp}$ in Fig.~\ref{Schematic_level})
and energy difference between the antibonding band of Cu $3d$ $x^2-y^2$ and O $2p_{\sigma}$ orbitals, and the Cu $3d$ $3z^2-r^2$ orbital hybridized mainly with the apex oxygen $p_z$ orbital ($\Delta E$ in Fig.~\ref{Schematic_level}).
\tred{When we successfully derive the Hamiltonians, it does not necessarily mean that the solutions of the Hamiltonians should appropriately describe the experimental results of the cuprate superconductors.  Instead, our Hamiltonians offer ways of understanding the validities of one-, two- and three-band Hamiltonians, and what the minimum effective Hamiltonians for the curates should be, for describing physics of the cuprates, which is still under extensive debate.}
}
\begin{widetext}
\begin{table*}[h!] 
\label{parameter_summary} 
\caption{
\tr{Summary of effective Hamiltonian parameters for HgBa$_2$CuO$_4$ and La$_2$CuO$_4$ (in eV).
$t$ and $t'$ for one- and two-band Hamiltonians are for nearest and next nearest neighbor transfers between Cu $3d$ orbitals, respectively.
Onsite and nearest neighbor interactions $U$ and $V$, respectively for Cu $3d$ orbitals are given as well. 
The orbital level is given by $\epsilon_{X}$ with $X=x^2-y^2$ or $3z^2-r^2$. Left panel:1-band Hamiltonians. Middle two panels: two-band Hamiltonians. 
Right panel: three-band Hamiltonians 
$t_{dp}$ ($t_{pp}$) is for largest nearest-neighbor transfer between Cu $3d_{x^2-y^2}$ and O $2p_{\sigma}$ (two O $2p_{\sigma}$) orbitals.
Onsite ($U$) and nearest neighbor ($V$) interactions for Cu $3d_{x^2-y^2}$ and O $2p_{\sigma}$ are given as well.
The level difference between $3d_{x^2-y^2}$ and $2p_{\sigma}$ is given by $\Delta_{dp}$. }
}
\begin{tabular}{cccc}
\begin{minipage}{0.18\hsize}
\label{summary1}
\begin{center}
\
\begin{tabular}{c|c} 
\hline \hline \\ [-8pt]
  HgBa$_2$CuO$_4$ & 1-band \\
\hline  \\ [-8pt]
$t$  &     -0.461    \\ [+1pt]
\hline \\ [-8pt] 
 $t'$    & 0.119    \\
\hline  \\ [-8pt]
$|t'/t|$  &  0.26     \\ [+1pt]
\hline \\ [-8pt] 
$U$  & 4.37 \\   [+1pt]
\hline \\ [-8pt] 
$V$  & 1.09 \\
\hline \\ [-8pt]  
 $|U/t|$   &  9.48    \\ 
\hline
\hline 
\\ [-8pt]
La$_2$CuO$_4$ & 1-band \\
\hline  \\ [-8pt]
$t$  &     -0.482    \\ [+1pt]
\hline \\ [-8pt] 
 $t'$    & 0.073    \\
\hline  \\ [-8pt]
$|t'/t|$  &  0.15     \\ [+1pt]
\hline \\ [-8pt] 
$U$  & 5.00 \\   [+1pt]
\hline \\ [-8pt] 
$V$  & 1.11 \\
\hline \\ [-8pt]  
 $|U/t|$   &  10.4   \\ \hline
\hline 
\end{tabular} 
\end{center} 
\end{minipage} 
\begin{minipage}{0.29\hsize}
\label{summary2}
\begin{center}
\
\begin{tabular}{c|c|c} 
\hline \hline \\ [-8pt]
HgBa$_2$CuO$_4$ & 2-band   \\ [+1pt]
\hline \\ [-8pt] 
  $t$    &  $3z^2-r^2 $ &  $x^2-y^2 $   \\ 
\hline \\ [-8pt] 
$3z^2-r^2 $  &  0.013 & 0.033  \\
$x^2-y^2 $  &  0.033  & -0.426  \\
\hline \\ [-8pt] 
  $t'$    &  $3z^2-r^2 $ &  $x^2-y^2 $  \\ 
\hline \\ [-8pt] 
$3z^2-r^2 $  &  -0.003 & 0.000  \\
$x^2-y^2 $  &  0.000  & 0.102  \\
\hline \\ [-8pt] 
  $|t_{x^2-y^2}'/t_{x^2-y^2}|$  & 0.24  \\ 
\hline \\ [-8pt] 
  $\epsilon_{x^2-y^2}-\epsilon_{3z^2-r^2}$ & \tg{4.01}  \\ 
\hline \\ [-8pt] 
  $U$    &  $3z^2-r^2 $ &  $x^2-y^2 $   \\ 
\hline \\ [-8pt] 
$3z^2-r^2 $  &  6.92 & 4.00 \\
$x^2-y^2 $  &  4.00  & 4.51  \\
\hline \\ [-8pt] 
  $V$    &  $3z^2-r^2 $ &  $x^2-y^2 $ \\ 
\hline \\ [-8pt] 
$3z^2-r^2 $  &  0.76 & 0.83  \\
$x^2-y^2 $  &  0.83  & 0.90  \\
\hline \\ [-8pt] 
  $|U/t_{x^2-y^2}|$    &  $3z^2-r^2 $ &  $x^2-y^2 $  \\ 
\hline \\ [-8pt] 
$3z^2-r^2 $  &  16.2 & 9.4  \\
$x^2-y^2 $  &  9.4  & 10.6 \\

\hline \hline 
\end{tabular} 
\end{center} 
\end{minipage} 
\begin{minipage}{0.29\hsize}
\label{summary2}
\begin{center}
\
\begin{tabular}{c|c|c} 
\hline \hline \\ [-8pt]
 La$_2$CuO$_4$ & 2-band  \\ [+1pt]
\hline \\ [-8pt] 
   $t$    &  $3z^2-r^2 $ &  $x^2-y^2 $  \\ 
\hline \\ [-8pt] 
$3z^2-r^2 $  & -0.008 & 0.057 \\
$x^2-y^2 $  & 0.057 & -0.389 \\
\hline \\ [-8pt] 
  $t'$    &  $3z^2-r^2 $ &  $x^2-y^2 $ \\ 
\hline \\ [-8pt] 
$3z^2-r^2 $  &   -0.013 & 0.000 \\
$x^2-y^2 $  &  0.000 & 0.136 \\
\hline \\ [-8pt] 
  $|t_{x^2-y^2}'/t_{x^2-y^2}|$   & 0.35 \\ 
[+1pt]
\hline \\ [-8pt] 
  $\epsilon_{x^2-y^2}-\epsilon_{3z^2-r^2}$ &   3.74 \\ 
\hline \\ [-8pt] 
  $U$    &  $3z^2-r^2 $ &  $x^2-y^2 $   \\ 
\hline \\ [-8pt] 
$3z^2-r^2 $   & 7.99 & 4.91\\
$x^2-y^2 $   & 4.91 & 5.48 \\
\hline \\ [-8pt] 
  $V$    &  $3z^2-r^2 $ &  $x^2-y^2 $   \\ 
\hline \\ [-8pt] 
$3z^2-r^2 $   &  1.43 &1.50 \\
$x^2-y^2 $  & 1.50 & 1.56 \\
\hline \\ [-8pt] 
  $|U/t_{x^2-y^2}|$    &  $3z^2-r^2 $ &  $x^2-y^2 $  \\ 
\hline \\ [-8pt] 
$3z^2-r^2 $  &   20.5 &12.6 \\
$x^2-y^2 $  &   12.6 & 11.6\\
\hline \hline 
\end{tabular} 
\end{center} 
\end{minipage} 
\begin{minipage}{0.22\hsize}
\label{summary3-1}
\begin{center}
\
\begin{tabular}{c|c} 
\hline \hline \\ [-8pt]
  HgBa$_2$CuO$_4$ & 3-band \\
\hline  \\ [-8pt]
$t_{dp}$  &     1.257    \\ [+1pt]
\hline \\ [-8pt] 
 $t_{pp}$    & 0.751    \\
\hline  \\ [-8pt]
$\Delta_{dp}$  &  \tg{2.416}     \\ [+1pt]
\hline \\ [-8pt] 
$U_{dd}$  & 8.84 \\   [+1pt]
\hline \\ [-8pt] 
$V_{dd}$  & 0.80 \\
\hline \\ [-8pt] 
$V_{dp}$  & 1.99 \\   [+1pt]
\hline \\ [-8pt] 
$U_{pp}$  & 5.31 \\
\hline \\ [-8pt]  
$V_{pp}$  & 1.21 \\
\hline \\ [-8pt]  
 $|U_{dd}/t_{dp}|$   &  7.03    \\ 
\hline
\hline 
\\ [-8pt]
La$_2$CuO$_4$ & 3-band \\
\hline  \\ [-8pt]
$t_{dp}$  &     \tg{1.369}    \\ [+1pt]
\hline \\ [-8pt] 
 $t_{pp}$    & \tg{0.754}    \\
\hline  \\ [-8pt]
$\Delta_{dp}$  &  \tg{3.699}     \\ [+1pt]
\hline \\ [-8pt] 
$U_{dd}$  & 9.61 \\   [+1pt]
\hline \\ [-8pt] 
$V_{dd}$  & 1.51 \\
\hline \\ [-8pt] 
$V_{dp}$  & 2.68 \\   [+1pt]
\hline \\ [-8pt] 
$U_{pp}$  & 6.13 \\
\hline \\ [-8pt]  
$V_{pp}$  & 1.86 \\
\hline \\ [-8pt]  
$|U_{dd}/t_{dp}|$   &  \tg{7.02}   \\ 
\hline
\hline 
\end{tabular} 
\end{center} 
\end{minipage} 
\end{tabular} 
\end{table*} 
\end{widetext}

\tred{In the present paper, we restrict the effective Hamiltonians into the standard form containing the kinetic and two-body interaction terms and ignore the multiparticle effective interactions more than the two-body terms.  This MACE scheme is based on the characteristic feature of strongly correlated electron systems, where the high-energy and low-energy degrees are  well separated and the partial trace out of the high-energy degrees of freedom can successfully be performed in perturbative ways as in the cRPA and cGW scheme\cite{ImadaMiyake,hirayama17}. In this perturbation expansion, the multiparticle effective interactions rather than the two-body terms are the higher order terms. Therefore, we ignore them in the same spirit with the cGW. }

In Sec. II we describe the basic method.  The three effective Hamiltonians for HgBa$_2$CuO$_4$ are derived in Sec. III.A and those for La$_2$CuO$_4$ are given in Sec.III.B. Section IV is devoted to discussions and we summarize the paper in Sec. V. 

\section{Method}
\subsection{Outline}
\subsubsection{Goal: Low-energy effective Hamiltonian}
Our goal of low-energy effective Hamiltonians for copper-oxide superconductors based on the cGW and SIC have the form
\begin{eqnarray}
\mathcal{H}_{\text{eff}} ^{\text{cGW-SIC}}= \sum_{ij} \sum_{\ell_1 \ell_2\sigma }&&
t^{\text{cGW-SIC}}_{\ell_1 \ell_2\sigma}(\bm{R}_i-\bm{R}_j) d_{i\ell_1\sigma} ^{\dagger} d_{j \ell_2\sigma} \nonumber \\
+ \frac{1}{2} \sum_{i_1i_2i_3i_4} \sum_{klmn \sigma \eta \rho \tau}  
&\biggl\{& W_{ \ell_1 \ell_2 \ell_3 \ell_4\sigma \eta \rho \tau }^r(\bm{R}_{i_1},\bm{R}_{i_2},\bm{R}_{i_3},\bm{R}_{i_4}) \nonumber \\
&&d_{i_1 \ell_1\sigma}^{\dagger}d_{i_2 \ell_2\eta} d_{i_3 \ell_3\rho}^{\dagger} d_{i_4 \ell_4\tau}\biggl\}.
\label{Hamiltonian0}
\end{eqnarray}
Here, the single particle term is represented by 
\begin{equation}
t^{\text{cGW-SIC}}_{ \ell_1 \ell_2\sigma}(\bm{R})= \langle \phi _{ \ell_1\bm{0}}|{H}^{\text{cGW-SIC}}_{K}|\phi _{ \ell_2\bm{R}} \rangle, 
\label{cGW-SICK}
\end{equation}
and  the interaction term is given by 
\begin{eqnarray}
W_{ \ell_1 \ell_2 \ell_3 \ell_4\sigma \eta \rho \tau }^r(\bm{R}_{i_1},\bm{R}_{i_2},\bm{R}_{i_3},\bm{R}_{i_4}) \nonumber \\
= \langle \phi _{ \ell_1\bm{R}_{i_1}}\phi _{ \ell_2\bm{R}_{i_2}}|{H}^{\text{cGW-SIC}}_{W^r}|\phi _{ \ell_3\bm{R}_{i_3}}\phi _{ \ell_4\bm{R}_{i_4}} \rangle, 
\label{cGW-SICW}
\end{eqnarray}
where ${H}^{\text{cGW-SIC}}={H}^{\text{cGW-SIC}}_{K}+{H}^{\text{cGW-SIC}}_{W^r}$ is the Hamiltonian in the continuum space obtained after the cGW and SIC treatments to the Kohn Sham (KS) Hamiltonian.
$t^{\text{cGW-SIC}}$  
represents transfer integral of the maximally localized Wannier functions (MLWF's)~\cite{marzari97,souza01} based on the cGW approximation supplemented by the SIC. Here, $\phi_{\ell\bm{R}}$ is the MLWF of the $\ell$th orbital localized at the unit cell $\bm{R}$.
We will show details of {the} cGW-SIC later.
Here, $d_{i\ell\sigma} ^{\dagger}$ ($d_{i \ell\sigma}$) 
is a creation (annihilation) operator of an 
electron with spin $\sigma$ in the $\ell$th MLWF centered at $\bm{R}_{i}$.

The dominant part of the screened interaction $W^r$ has the form 
\begin{eqnarray} 
 U_{ \ell_1\ell_2\sigma \rho }(\bm{R}_i-\bm{R}_j) 
&=& W_{ \ell_1 \ell_1 \ell_2 \ell_2\sigma \sigma \rho \rho }^r
(\bm{R}_i,\bm{R}_i,\bm{R}_j,\bm{R}_j) \label{Hamiltonian1}  
\end{eqnarray}
for the diagonal interaction including the onsite intraorbital term $U_{ \ell}=U_{ \ell \ell\sigma -\sigma }(\bm{R}_i-\bm{R}_j=0) $ and the spin-independent onsite interorbital terms  $U_{\ell_1 \ell_2}'=U_{\ell_1 \ell_2\sigma \rho }(\bm{R}_i-\bm{R}_j=0) $ (for $ \ell_1\ne  \ell_2$) as well as spin-independent intersite terms
$V_{ij  \ell_1\ell_2}=U_{\ell_1\ell_2\sigma \rho }(\bm{R}_i-\bm{R}_j) $,
where we assume the translational invariance.
In addition, the exchange terms 
\begin{eqnarray} 
J_{\ell_1 \ell_2\sigma \rho}(\bm{R}_i-\bm{R}_j) 
&=&W_{\ell_1\ell_1 \ell_2 \ell_2\sigma \rho \rho \sigma }^r(\bm{R}_{i},\bm{R}_{i},\bm{R}_{j},\bm{R}_{j})
\nonumber \\
&=&W_{ \ell_1 \ell_2 \ell_1 \ell_2\sigma \rho \rho \sigma }^r(\bm{R}_{i},\bm{R}_{j},\bm{R}_{i},\bm{R}_{j})
\label{Hamiltonian2}  
\end{eqnarray} 
have nonnegligible contributions, particularly for the onsite tems where $\bm{R}_{i}=\bm{R}_{j}$. Other off-diagonal terms are in general smaller than 50 meV in our result of the cuprate superconductors and mostly negligible. 

\subsubsection{Basic downfolding scheme}
We start from the conventional local density approximation (LDA) for the global band structure, which is justified because strong correlation effects and quantum fluctuations far from the Fermi level are weak.  For the central part near the Fermi level, we consider later beyond LDA.  
Our LDA calculation is based on the full potential linearized muffin tin orbital (FP-LMTO) method\cite{AndersenLMTO}. 

To remove the double counting of the Coulomb exchange contributions,
we completely subtract the exchange correlation contained in the LDA calculation and replace it with the cGW calculation, where the self-energy effects are taken into account only for those containing the contribution from outside of the target low-energy effective Hamiltonian, because the self-energy in the effective Hamiltonian will be considered later by more refined methods beyond GW. 

More specifically, since we derive three effective Hamiltonians, we employ two steps for an efficient derivation. First we derive the effective Hamiltonians for 17 bands near the Fermi level whose main components are from 5 Cu $3d$ orbitals, and 3 oxygen $2p_{\sigma}$ orbitals at 2 O atoms each in the CuO$_2$ plane and at 2 other out-of-plane O atoms each above and below Cu in a unit cell. In fact, the 17 bands near the Fermi level are relatively well separated from other high-energy bands (namely, bands far from the Fermi level) and the 17 bands Hamiltonians offer a good base for the next step. Then thanks to the chain rule~\cite{aryasetiawan04,ImadaMiyake}, we derive three different types of effective Hamiltonians successively from the 17-band  effective Hamiltonian. We abbreviate the electronic degrees of freedom outside the 17 bands as H and those of 17 bands M which excludes the final target space L for the low-energy effective Hamiltonian.  We also employ the abbreviation N for the electronic degrees of freedom consisting of both of L and M.
\tg{The hierarchical structure described above is shown in Fig.~\ref{Schematic_Screening}}
\begin{figure}[ptb]
\centering 
\includegraphics[clip,width=0.3\textwidth ]{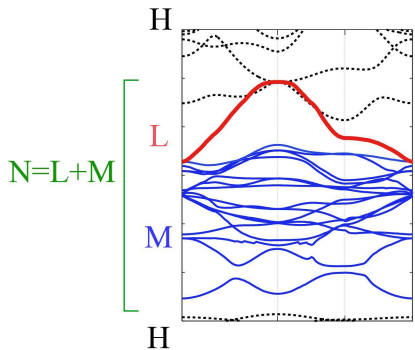} 
\caption{(Color online) Hierarchical structure in the procedure of the downfolding. The black dashed bands H in the high energy part are first downfolded to the renormalized 17 bands described by N. Then the M bands (blue thin bands) among N are eliminated and renormalized into the final low-energy effective Hamiltonian constructed from L (red thick bands). Here, an example of the procedure to derive a one-band Hamiltonian is shown. }
\label{Schematic_Screening}
\end{figure} 

\vspace{2mm}
\noindent
\tc{{\it -- From full Hilbert space to 17-band subspace --}} \\
Let us first describe the first cGW scheme\cite{hirayama13,hirayama17} to derive the 17-band effective Hamiltonian for N near the Fermi level.  
After removing the exchange correlation potential contained in the LDA calculation, 
we first perform the full GW calculation for the 17 bands. 
This GW scheme allows to completely remove the double counting of the correlation effect arising from the exchange correlation energy in LDA. 
Here, the full GW calculation is defined as that takes into account the self-energy effect calculated using the fully screened interaction $W$ including the screening by electrons in all the bands.  The reason why we use the full GW is based on the spirit that the screening from the 17 bands taken into account later on are better counted by using its renormalized level.   
    
In the present work, except La $4f$ band in La$_2$CuO$_4$, we retain the LDA dispersion for the bands other than the 17 bands, because their renormalization have few effects on the final low-energy effective Hamiltonian. For La $4f$ band in La$_2$CuO$_4$, it is known that the LDA calculation \tr{qualitatively fails in counting its correlation effects and the insulating nature~\cite{Mattheiss1987,Massidda1987,Pickett1991}, which is also related to the fact that the LDA incorrectly gives the level too close to the Fermi level~\cite{fujimori1987}}. Then we first perform the one-shot GW calculation for the La $4f$ band before the full GW calculation for the 17 bands.

We then perform the cGW calculation for the 17 bands, where the self-energy is calculated from the full GW Green's function $G^{(\rm GW)}$ for the 17 bands and the LDA Green's function for the other high-energy bands. 
After disentanglement between the H and N bands by the conventional method\cite{miyake09}, we assume that the non-interacting Green's function $G^{(\rm GW)}$ is 
block-diagonal and can be decomposed into
\begin{eqnarray}
G^{(\rm GW)} &=& G^{(\rm GW)}_{ll} |L \rangle \langle L |+G^{(\rm GW)}_{mm} |M \rangle \langle M | \nonumber \\
&+& G^{(\rm LDA)}_{hh} |H \rangle \langle H |
\end{eqnarray}
where $|H \rangle, |M \rangle$ and $|L \rangle$ represent the respective subspaces.  
We use the notation $G_{ab} = - \langle T c_{a}(\tau) c^{\dagger}_{b} \rangle$,
where $a, b$ denote elements either $h,m$ or $l$. Here, $h$, $m$ and $l$ represent 
bands belonging to H, M and L degrees of freedom, respectively. We also introduce $W_{abcd}$ 
for the coefficient of the interaction term $c^{\dagger}_ac_bc^{\dagger}_cc_d$. 
We calculate the partially screened Coulomb interaction $W_{\rm N}$ that contains only the screening contributed from the H space\cite{hirayama13,hirayama17}.

Then with the notation $|N \rangle$ ($n$) for the subspace containing $|L \rangle$ and $|M \rangle$ ($l$ and $m$) together, the constrained self-energy at this stage, $\Sigma_{\rm H}$ is described from the full GW self-energy 
\begin{eqnarray}
 \Sigma =\Sigma_{nn}+\Sigma_{nh}G_{hh}\Sigma_{hn} ,
\label{Sdd1}
\end{eqnarray}
where
\begin{eqnarray}
\Sigma_{nh}(q,\omega)
&=& [G^{(\rm GW)}_{nn} W_{nnnh}](q,\omega) \nonumber \\
                                                      &+& G^{(\rm LDA)}_{hh} W_{nhhh}(q,\omega)
\label{Sdr}
\\  
\Sigma_{nn}(q,\omega)&=&  [G^{(\rm GW)}_{nn} W_{nnnn}](q,\omega) \nonumber \\ 
                                                    &+& [G^{(\rm LDA)}_{hh} W_{nhhn}](q,\omega) 
\label{Sdd2}
\end{eqnarray}
as
\begin{eqnarray}
\Sigma_{{\rm H}{nn'}}(q,\omega)
&=& \Sigma_{nn'}(q,\omega) \nonumber \\
&-& \sum_{n_1,n_2} [G^{(\rm GW)}_{n_1n_2} W_{nn_1n_2n'}](q,\omega).
\label{Sdd4}
\end{eqnarray}
In Eqs.(\ref{Sdr}) and (\ref{Sdd2}), the right hand side terms are the only nonzero terms because $G$ is assumed that it does not have off-diagonal element between N and H. \tred{The off-diagonal part can be ignored because they are higher-order terms in the GW scheme (see also the reason for ignoring the off-diagonal part)~\cite{hirayama17}.}
Here the notation $[GW](q,\omega)$ represents the convolution
\begin{eqnarray}
 [GW](q,\omega )= \int d\omega' dq'  G(q',\omega' )W(q+q',\omega+\omega' ).
\label{Sdd3}
\end{eqnarray}
In the present study, we neglect the second term in the right hand side of Eq.(\ref{Sdd1}) because it is small higher-order term. The first term in  Eq.(\ref{Sdd2}) is excluded to avoid double counting because this is the term to be considered in the 
low-energy solver.

\tred{If one wishes to construct a low-energy Hamiltonian by reducing to the static effective interaction,} this constrained self-energy $\Sigma_{\rm H}(q,\omega)$ is supplemented by the constrained self-energy $\Sigma_{\rm H}^{\rm dyn}(q,\omega)$ arising from the frequency-dependent part of the  screened interaction\cite{hirayama13,hirayama15,hirayama17}
described by 
\begin{eqnarray}
\Sigma_{\rm H}^{\rm dyn}
&=& 
G_{nn}^{(\rm GW)} W^{\rm dyn}_{\rm N}.
\label{DeltaSigLnonlocalGW}
\end{eqnarray}
Here, 
$W^{\rm dyn}_{\rm N}$ is defined by 
\begin{eqnarray}
W^{\rm dyn}_{\rm N}(q,\omega)\equiv W(q,\omega) - W_{\rm N} (q,\omega),
\label{WdynGW}
\end{eqnarray}
where $W$ is the fully screened interaction in the RPA level as
\begin{eqnarray}
W (q,\omega) = \frac{v(q)}{1- P(q,\omega)v(q)}.
\label{WdynGW2}
\end{eqnarray}
$W_{\rm N} (q,\omega)$ is the ``fully screened interaction" within the N space; 
\begin{eqnarray}
W_{\rm N} (q,\omega) = \frac{W_{\rm H}(q, \omega=0)}{1- P_{\rm N}(q,\omega)W_{\rm H}(q, \omega=0)},
\label{WdynGW3}
\end{eqnarray}
\tred{(If one solves the frequency dependent effective interaction as it is in the Lagrangian form, this procedure is not necessary.)}
Here, $W_{\rm H}$ is the partially screened interaction obtained 
from the cRPA in the spirit of excluding the polarization within the 17 bands.
Namely,
\begin{eqnarray}
W_{\rm H} (q,\omega) = \frac{v(q)}{1- P_{\rm H}(q,\omega)v(q)},
\label{Wr_crpa}
\end{eqnarray}
where the wave-number ($q$) dependent bare Coulomb interaction $v$ 
is partially screened by the partial polarization $P_{\rm H}$. 
Here, $P_{\rm H}$ is defined in terms of the total polarization $P$ by excluding 
the intra-N-space polarization $P_{\rm N}$: $P_{\rm H}\equiv P-P_{\rm N}$.
$P_{\rm N}$ involves only screening processes within the N-space. 
Namely, in the cRPA, the polarization without low-energy N-N transition $P_{\rm H}$ are estimated as,
\begin{equation}
-P_{\rm H}=iGG-iG_{\rm N}G_{\rm N}=iG_{\rm N}G_{\rm H}+iG_{\rm H}G_{\rm N}+iG_{\rm H}G_{\rm H},
\label{PH}
\end{equation}
where the whole Green's function $G$ is given by the sum of the low- and high-energy propagators estimated by the GW for $G_{\rm N}$ and by the LDA for $G_{\rm H}$, respectively. 
Then in Eq.(\ref{WdynGW3}), $W_{\rm H}(q, \omega=0)$ plays the role of ``bare interaction" within the N space.
Eventually, $W_{\rm N}^{\rm dyn}$ is
the frequency-dependent part of the interaction 
that would be missing if the 17-band N part were solved
within the GW approximation. (See \tb{the horizontal-stripped area in} Fig.~\ref{Schematic_Screening}, see also  Fig.1 in Ref.~\onlinecite{hirayama17}).

\tb{Here, we note that, instead of the dynamical part $W_{\rm N}^{\rm dyn}$ in Eq. (\ref{WdynGW}), we could take $W_{\rm H}(q,\omega)-W_{\rm H}(q,\omega=0)$ as a naive choice of the dynamical part, which is depicted as the vertical-stripped area.
However, Eq. (\ref{WdynGW}) is expected to express the dynamical part more
accurately because Eq. (\ref{WdynGW}) takes into account the RPA level fluctuations (though not perfect) beyond $W_{\rm H}(q,\omega)-W_{\rm H}(q,\omega=0)$.
First, we \tc{note} that the \tc{interaction part of effective Hamiltonians we derive must be expressed in the form of screened but static Coulomb interactions.}
Therefore, the dynamical part of the Coulomb interactions due to the screening from the high-energy degrees of freedom is taken into account as the self-energy correction.
\tc{Now}, $W$ is the \tc{fully screened} dynamical interaction in the RPA level and $W_{\rm N}$ is the screened interaction if the effective Hamiltonian with the static interaction $W_{\rm H}(q,\omega=0)$ would be solved in the same RPA level.
Then, the difference between $W$ and $W_{\rm N}$, which is nothing but $W_{\rm N}^{\rm dyn}$, \tc{is the part we ignore when we solve the effective Hamiltonian with with the static interaction $W_{\rm H}(q,\omega=0)$ by the RPA. Therefore, $W_{\rm N}^{\rm dyn}$} should be taken into account as the self-energy correction in the present downfoldin scheme.]
}
\begin{figure}[ptb]
\centering 
\includegraphics[clip,width=0.3\textwidth ]{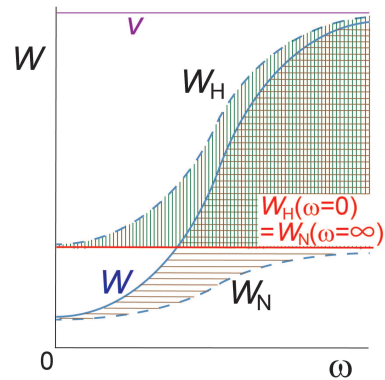} 
\caption{(Color online) Schematic frequency dependence of
effective interaction screened from bare interaction $v$. Other interactions are obtained from full RPA (GW) ($W$), cRPA ($W_{\rm H}$) and screened
interaction by RPA ($W_{\rm N}$) within low-energy effective Hamiltonian
at the effective interaction $W_{\rm H}(\omega = 0)$. \tb{The vertical stripped area represents the dynamical part of cRPA-screened interaction $W_{\rm H}$, which is not contained in the effective Hamiltonian with the static interaction $W_{\rm H}(\omega = 0)$. This part requires additional treatments. Instead of the vertical-stripped area, the horizontal-stripped area, $W-W_{\rm N}$ (Eq.(\ref{WdynGW})), can be regarded as a better choice for the dynamical part to be treated additionally (see the text).}}
\label{Schematic_Screening}
\end{figure} 

Thus, the constrained renormalized Green's function for the 17-band effective Hamiltonian is described by   
\begin{eqnarray} 
&&G_{\rm N}(\omega )
=\frac{I}{\omega I-(H^{\text{LDA}}-V^{\text{xc}}+\Sigma_{\rm H}+\Sigma_{\rm H}^{\rm dyn})}  \nonumber \\
&\approx & \frac{Z_{\rm H}(\epsilon^{\text{GW}})}{\omega I-(H^{\text{LDA}}+ 
 Z_{\rm H}^{\rm cGW}(\epsilon^{\text{GW}})(-V^{\text{xc}}+(\Sigma_{\rm H}+\Sigma_{\rm H}^{\rm dyn})(\epsilon^{\text{GW}})))}, \nonumber \\
&\approx & \frac{I}{\omega I-Z_{\rm H}^{\rm cGW}(0)(H^{\text{LDA}}-V^{\text{xc}}+\text{Re}(\Sigma _{\rm H}+ \Sigma_{\rm H}^{\rm dyn})(0))},
\label{rdelG}
\end{eqnarray}
\begin{equation}
Z_{\rm H}^{\rm cGW}(\epsilon)= \biggl\{ I-\frac{\partial (\text{Re}\Sigma_{\rm H}+\text{Re}\Sigma_{\rm H}^{\rm dyn})}{\partial \omega }\Big|_{\omega =\epsilon } \biggr\}^{-1},
\label{Zrdel}
\end{equation}
where we have suppressed writing the explicit wavenumber and orbital dependence and $\epsilon^{\rm GW}$ is the band energy by the GW calculation.
Then the one-body part of the static effective Hamiltonian for the 17 bands in the cGW is given by~\cite{hirayama13}  
\begin{eqnarray}
{H}^{\text{cGW}}_{{\rm N}K}&=&\sum_{n_1n_2} \tilde{H}^{\text{cGW}}_{{\rm N}Kn_1n_2} \nonumber \\
\tilde{H}^{\text{cGW}}_{{\rm N}Kn_1n_2} &=&\sum_qZ_{{\rm H}n_1n_2}^{\rm cGW}(q,\epsilon=0)
{H}^{\text{cGW-H}}_{{\rm N}K}(q)
\\
\tilde{H}^{\text{cGW-H}}_{{\rm N}Kn_1n_2}(q)&=&H^{\text{LDA}}_n(q)\delta_{n_1n_2}-V^{\text{xc}}n_1n_2(q) \nonumber \\
&&+\text{Re}(\Sigma _{{\rm H}n_1n_2}+ \Sigma_{{\rm H}n_1n_2}^{\rm dyn})(q,\omega=0),
\label{HeffcGW-SIC}
\end{eqnarray}
which is represented by the first quantization form in the continuum space. 

The effective interactions for the 17 bands have also been calculated by using cRPA~\cite{aryasetiawan04,ImadaMiyake}, 
where effects of polarization contributing from the other bands are taken into account as a partially screened interaction. 
The partially screened Coulomb interaction for the 17 bands is given by
\begin{eqnarray}
&&W_{{\rm N}n_1n_2n_3n_4\sigma \eta \rho \tau }(\bm{R}_{i_1},\bm{R}_{i_2},\bm{R}_{i_3},\bm{R}_{i_4}) \nonumber \\
&=& \langle \phi^{\rm N}_{n_1\bm{R}_{i_1}}\phi^{\rm N}_{n_2\bm{R}_{i_2}}|W_{\rm N}(\omega=\infty)|\phi^{\rm N}_{n_3\bm{R}_{i_3}}\phi^{\rm N}_{n_4\bm{R}_{i_4}}\rangle ,
\label{W_N}
\end{eqnarray}
where $\phi^{\rm N}_{n\bm{R}}$ represents the MLWF for the 17 bands (the orbital index $n$ runs from 1 to 17). Note that ${W}_{\rm N}(\omega=\infty)$ is nothing but ${W}_{\rm H}(\omega=0)$(see Fig.~\ref{Schematic_Screening}).


Then the 17-band cGW effective Hamiltonian for the lattice fermions in the second-quantized Wannier orbitals representation is given by 
\begin{eqnarray}
&&\mathcal{H}_{\text{N}} ^{\text{cGW}}= \mathcal{H}_{\text{N}K} ^{\text{cGW}}+\mathcal{H}_{\text{N}W} ^{\text{cGW}} \label{HNCGW} \\
&&\mathcal{H}_{\text{N}K} ^{\text{cGW}}=\sum_{ij} \sum_{n_1n_2\sigma }
t^{\text{cGW}}_{{\rm N}n_1n_2\sigma}(\bm{R}_i-\bm{R}_j) d_{in_1\sigma} ^{\dagger} d_{jn_2\sigma} 
\label{HNKCGW} \\
&&\mathcal{H}_{\text{N}W} ^{\text{cGW}} = \nonumber \\
&&\frac{1}{2} \sum_{i_1i_2i_3i_4} \sum_{n_1n_2n_3n_4 \sigma \eta \rho \tau}  
\biggl\{ W_{{\rm N}n_1n_2n_3n_4\sigma \eta \rho \tau }(\bm{R}_{i_1},\bm{R}_{i_2},\bm{R}_{i_3},\bm{R}_{i_4}) \nonumber \\
&&
d_{i_1n_1\sigma}^{\dagger}d_{i_2n_2\eta} d_{i_3n_3\rho}^{\dagger} d_{i_4n_4\tau}\biggl\}.
\label{HNWCGW}
\end{eqnarray}
Here, the single particle term is represented by 
\begin{equation}
t^{\text{cGW}}_{{\rm N}n_1n_2\sigma}(\bm{R})= \langle \phi _{n_1\bm{0}}|{H}^{\text{cGW}}_{{\rm N}Kn_1n_2}|\phi _{n_2\bm{R}} \rangle, 
\label{cGW-SICK_N}
\end{equation}

In addition, we supplement in the single-particle term, 
the self-interaction correction (SIC) to recover the cancellation realized in LDA. Since we subtracted the exchange correlation energy, the cancellation with the counterpart of the Hartree term becomes violated. To recover the cancellation, we impose the correction following Ref.\onlinecite{hirayama13}. 
The SIC in  
the 17-band degrees of freedom is $U^{\text{on-site}}_{{\rm N}n} n_{{\rm N}n\text{GW}}/2 $  
where $U^{\text{on-site}}_{{\rm N}n}=W_{{\rm N}nnnn \sigma \sigma -\sigma -\sigma }(\bm{R},\bm{R},\bm{R},\bm{R})$ is the on-site effective interaction for the band $n$
and $n_{{\rm N}n \text{GW}}$ is the occupation number of the $n$-th band 
for the 17 bands including up and down spins in the GW calculation.
Then the cGW-SIC effective Hamiltonian for the 17 bands is given by
\begin{eqnarray}
\mathcal{H}^{\text{cGW-SIC}}_{\rm N}&=&\mathcal{H}^{\text{cGW-SIC}}_{{\rm N}K}+\mathcal{H}^{\text{cGW}}_{{\rm N}W} \\
\mathcal{H}^{\text{cGW-SIC}}_{{\rm N}K}&=&\mathcal{H}^{\text{cGW}}_{{\rm N}K} \nonumber \\
-\sum_{in\sigma}&Z_{{\rm H}n}^{\text{cGW}}&(q=\omega=0)U^{\text{on-site}}_n \frac{d_{in\sigma}^{\dagger} d_{in\sigma}}{2}  
\label{HcGW-SIC}
\end{eqnarray}
The renormalization factor $Z_{{\rm H}n}^{\text{cGW}}$ is needed to renormalize 
the frequency-dependent part of the interaction into a static effective Hamiltonian~\cite{hirayama13}.

An advantage of the MACE downfolding scheme in the procedure of deriving low-energy effective Hamiltonian is that the degrees of freedom retained in the low-energy effective Hamiltonians for the 
electrons near the Fermi level (electrons in the target bands) can be reduced progressively from the effective Hamiltonian containing larger number of bands to smaller, thanks to the chain rule of the cRPA in a controlled manner\cite{ImadaMiyake}.

By using this sequential downfolding scheme, we derive three types of effective Hamiltonians from the 17-bands effective Hamiltonians for the two compounds.  
The three types are for the electrons mainly originated from \\
1) the antibonding orbital generated from Cu 3d $x^2-y^2$ orbitals strongly hybridized with O $2p_{\sigma}$ orbitals (one-band effective Hamiltonian) \\ 
2) the antibonding orbital in 1) together with Cu $3d$ $3z^2-r^2$ orbital hybridized with the apex oxygen $p_z$ orbital (two-band effective Hamiltonian) \\ 
3) Cu 3d $x^2-y^2$ orbitals and two O $2p_{\sigma}$ orbitals aligned in the direction to Cu (three-band effective Hamiltonian).  \\ 
The degrees of freedom (bands) contained in these final Hamiltonians are called the target degrees of freedom (target bands). \tg{Although it is possible to derive Hamiltonians 
consisting of more than three bands such as four- or
six-orbital Hamiltonians, additional orbitals are fully \tp{occupied} even after the correlation effects are taken into account and expected to play minor role in the low-energy
physics. Thus, we mainly consider the above three types
of low-energy effective Hamiltonians.}

\vspace{2mm}
\noindent
\tc{{\it -- From 17-band subspace to low-energy effective Hamiltonians --}} \\
After restricting the Hilbert space to the 17-band Hamiltonian, 
we again employ the cGW scheme~\cite{hirayama13,hirayama17,aryasetiawan09} that 
additionally accounts for the self-energy within the 17-band Hilbert space. However, we exclude  that arising solely from the target bands to remove the double counting because it will be counted when the effective Hamiltonian is solved afterwards.
In this cGW scheme, the energy levels of the 17 bands are given from the former cGW level given in Eq.(\ref{HNKCGW}) as the starting point. 
Through the cGW scheme, the fully screened interaction is again employed in the calculation of the self-energy.
The constrained self-energy of the target band is further improved by considering the renormalization effect from the frequency dependent part of the effective interaction based on the cGW scheme in the same way as before\cite{hirayama13,hirayama17}.  

This two-step procedure is equivalent to the single procedure to directly derive the three Hamiltonian. In this second step, we restrict the electronic Hilbert space into the N space.  Then one simply needs to replace H with M, N with L and $v$ with $W_{\rm H}(\omega=0)$ in the procedure from Eq.(\ref{Sdd1}) to (\ref{Wr_crpa})(In Fig.\ref{Schematic_Screening}, $v, W_{\rm H}$ and $W_{\rm N}$ should be replaced with $W_{\rm H}(\omega=0), W_{\rm M}$ and $W_{\rm L}$, respectively.)

More concretely, the low-energy Hamiltonian includes the self-energy effects from the M degrees of freedom similarly to Eq.(\ref{HeffcGW-SIC}) as 
\begin{eqnarray}
{H}^{\text{cGW}}_{{\rm L}K}&=&\sum_{l_1l_2} \tilde{H}^{\text{cGW}}_{{\rm L}Kl_1l_2} \nonumber \\
\tilde{H}^{\text{cGW}}_{{\rm L}Kl_1l_2} &=&\sum_q Z_{{\rm HM}l_1l_2}^{\rm cGW}(q,\epsilon=0 ) \nonumber \\
\times (\tilde{H}^{\text{cGW-H}}_{{\rm N}Kl_1l_2}&(q)& +\text{Re}(\Sigma _{{\rm M}l_1l_2}+ \Sigma_{{\rm M}l_1l_2}^{\rm dyn})(q,\omega=0 )),
\label{HeffLcGW-SIC}
\end{eqnarray}
where $\Sigma _{{\rm M}l_1l_2}$ is the constrained self-energy that excludes that arising from the L degrees of freedom.
Namely, we utilize
\begin{eqnarray}
 \Sigma_{{\rm N}l_1l_2} =\Sigma_{l_1l_2}+\sum_{m_1,m_2}\Sigma_{l_1m_1}G_{m_1m_2}\Sigma_{m_2l_2} ,
\label{SddM1}
\end{eqnarray}
with
\begin{eqnarray}
\Sigma_{lm}(q,\omega)
&=&\sum_{l_1l_2}[ G^{(\rm GW)}_{l_1l_2} W_{ll_1l_2m}](q,\omega) \nonumber \\
                                                      &+& \sum_{m_1m_2}[G^{(\rm GW)}_{m_1m_2} W_{lm_1m_2m}](q,\omega)
\label{SdrL}
\\  
\Sigma_{ll}(q,\omega)&=&  \sum_{l_1l_2}[G^{(\rm GW)}_{l_1l_2} W_{ll_1l_2l}](q,\omega) \nonumber \\ 
                                                    &+&\sum_{m_1m_2} [G^{(\rm GW)}_{m_1m_2} W_{lm_1m_2l}](q,\omega),
\label{SddM2L}
\end{eqnarray}
where $l$ in $\Sigma_{ll}$ of Eq.(\ref{SddM2L}) represents inclusive terms containing the off-diagonal elements within the L space as in Eqs.(\ref{Sdd1}), (\ref{Sdr}) and (\ref{Sdd2}).  
Then, in contrast to Eq.(\ref{Sdd1}), we take into account the second term in Eq. (\ref{SddM1}) but similarly exclude the first term in Eq.(\ref{SddM2L}). 
Then $\Sigma _{{\rm M}ll'}$ is given by
\begin{equation}
 \Sigma_{{\rm M}{ll'}}(q,\omega) =\Sigma_{{\rm N}{ll'}}(q,\omega)-\sum_{l_1l_2}[G^{(\rm GW)}_{l_1l_2} W_{ll_1l_2l'}](q,\omega),
\label{SddM3}
\end{equation}

The renormalization factor in Eq.(\ref{HeffLcGW-SIC}) is given by
\begin{eqnarray}
&&Z_{\rm HM}^{\rm cGW}(\epsilon) \nonumber \\
&&= \biggl\{ I-\frac{\partial (\text{Re}\Sigma_{\rm H}+\text{Re}\Sigma_{\rm H}^{\rm dyn}+\text{Re}\Sigma_{\rm M}+\text{Re}\Sigma_{\rm M}^{\rm dyn})}{\partial \omega }\Big|_{\omega =\epsilon } \biggr\}^{-1}.
\label{Zrdel}
\end{eqnarray}

\begin{figure}[ptb]
\centering 
\includegraphics[clip,width=0.3\textwidth ]{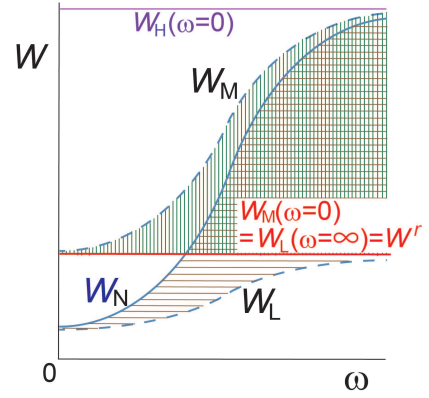} 
\caption{(Color online) Schematic frequency dependence of
effective interaction screened within the 17 band. Within the 17 bands, $W_{\rm H}(\omega=0)$ plays the role of the bare interaction and other interactions are obtained from full RPA (GW) ($W_{\rm N}$), cRPA ($W_{\rm M}$) and screened
interaction by RPA ($W_{\rm L}$) within low-energy effective Hamiltonian
at the effective interaction $W_{\rm M}(\omega = 0)$. \tb{The vertical and horizontal stripped area have similar meanings to those in Fig.~\ref{Schematic_Screening}}}
\label{Schematic_Screening_L}
\end{figure}

In the same way as Eqs.~(\ref{DeltaSigLnonlocalGW}) and (\ref{WdynGW}),
we use the following relations:
\begin{eqnarray}
\Sigma_{\rm M}^{\rm dyn}
&=& 
G_{ll}^{(\rm GW)} W^{\rm dyn}_{\rm L}.
\label{DeltaSigLnonlocalGW2L}
\end{eqnarray}
Here, $W^{\rm dyn}_{\rm L}$ is defined by
\begin{eqnarray}
W^{\rm dyn}_{\rm L}(q,\omega)\equiv W_{\rm N}(q,\omega) - W_{\rm L} (q,\omega).
\label{WdynGW2L}
\end{eqnarray}
(See \tb{the horizontal-stripped area} Fig.\ref{Schematic_Screening_L}).

The single-particle term is then 
\begin{eqnarray}
&&\mathcal{H}_{\text{L}K} ^{\text{cGW}}=\sum_{ij} \sum_{l_1l_2\sigma }
t^{\text{cGW}}_{{\rm L}l_1l_2\sigma}(\bm{R}_i-\bm{R}_j) d_{il_1\sigma} ^{\dagger} d_{jl_2\sigma} 
\label{HLKCGW},
\end{eqnarray}
where by using Eq.(\ref{HeffLcGW-SIC}),
\begin{equation}
t^{\text{cGW}}_{{\rm L}l_1l_2\sigma}(\bm{R})= \langle \phi _{l_1\bm{0}}^{\rm L}|{H}^{\text{cGW}}_{{\rm L}K}|\phi _{l_2\bm{R}}^{\rm L} \rangle, 
\label{cGW_L}
\end{equation}
has the form in Eq.(\ref{cGW-SICK}).

We also consider the self-interaction correction as
\begin{eqnarray}
\mathcal{H}^{\text{cGW-SIC}}_{{\rm L}K}&=&\mathcal{H}^{\text{cGW}}_{{\rm L}K} -\sum_{il\sigma}Z_{{\rm HM}l}^{\text{cGW}}(q=\epsilon=0)\nonumber \\
&\times&U^{\text{on-site}}_l \frac{d_{il\sigma}^{\dagger} d_{il\sigma}}{2}  
\label{HcGW-SIC_L}
\end{eqnarray}
The renormalization factor $Z_{{\rm HM}l}^{\text{cGW}}$ is again needed to renormalize 
the frequency-dependent part of the interaction into a static effective Hamiltonian~\cite{hirayama13}.
Here, $U^{\text{on-site}}_{{\rm L}l}=W_{{\rm L}llll \sigma \sigma -\sigma -\sigma }(\bm{R},\bm{R},\bm{R},\bm{R})$ is the on-site effective interaction for the band $l$.

For the interaction parameter of the target effective Hamiltonian $W_{{\rm L}l_1l_2l_3l_4\sigma \eta \rho \tau }(\bm{R}_{i_1},\bm{R}_{i_2},\bm{R}_{i_3},\bm{R}_{i_4})$,
we apply the cRPA again now within the 17 band Hamiltonian.  Our task here is the procedure similar to that from Eqs.(\ref{WdynGW2}) to (\ref{Wr_crpa}), but replace H and N with M and L, respectively, where L represents the target bands. Thanks to the chain rule, this derivation of the effective interaction looks the same as the direct single step cRPA for the whole bands.  However, since the energy levels are replaced with the full GW energy levels within the 17 bands, the effective interaction is more refined by taking into account the self-energy effect for the 17 bands.  

Then 
\begin{eqnarray}
W_{\rm M} (q,\omega) &=& \frac{W_{\rm H}(q, \omega=0)}{1- P_{\rm M}(q,\omega)W_{\rm H}(q, \omega=0)},
\label{WdynGW4} \\
W_{\rm L} (q,\omega) &=& \frac{W_{\rm M}(q, \omega=0)}{1- P_{\rm L}(q,\omega)W_{\rm M}(q, \omega=0)},
\label{WdynGW5}
\end{eqnarray}
are satisfied within the N Hilbert space.  

Now the goal of our low-energy cGW effective Hamiltonian  is given by 
\begin{eqnarray}
&&\mathcal{H}_{\text{L}} ^{\text{cGW-SIC}}= \mathcal{H}_{\text{L}K} ^{\text{cGW-SIC}}+\mathcal{H}_{\text{L}W} ^{\text{cGW}} \label{HLCGW} \\
&&\mathcal{H}_{\text{L}W} ^{\text{cGW}} = \nonumber \\
&&\frac{1}{2} \sum_{i_1i_2i_3i_4} \sum_{l_1l_2l_3l_4 \sigma \eta \rho \tau}  
\biggl\{ W^r_{l_1l_2l_3l_4\sigma \eta \rho \tau }(\bm{R}_{i_1},\bm{R}_{i_2},\bm{R}_{i_3},\bm{R}_{i_4}) \nonumber \\
&&d_{i_1l_1\sigma}^{\dagger}d_{i_2l_2\eta} d_{i_3l_3\rho}^{\dagger} d_{i_4l_4\tau}\biggl\},
\label{HLWCGW}
\end{eqnarray}
where the single particle term $\mathcal{H}_{\text{L}K} ^{\text{cGW-SIC}}$ is given by Eqs.(\ref{HLKCGW}) and  (\ref{HcGW-SIC_L})
in the form Eq.(\ref{cGW-SICK}) and the interaction term has the form (\ref{cGW-SICW}) given by
\begin{eqnarray}
&W&_{ l_1 l_2 l_3 l_4\sigma \eta \rho \tau }^r(\bm{R}_{i_1},\bm{R}_{i_2},\bm{R}_{i_3},\bm{R}_{i_4}) \nonumber \\
&=& \langle \phi _{ l_1\bm{R}_{i_1}}\phi _{ l_2\bm{R}_{i_2}}|W_{\rm L}(\omega=\infty)|\phi _{ l_3\bm{R}_{i_3}}\phi _{ l_4\bm{R}_{i_4}} \rangle
\label{cGW-SICW_L}
\end{eqnarray}

\tred{If one wishes to solve the low-energy Hamiltonian by the dynamical mean-field approximation, the nonlocal part of the interaction is hardly taken into account. The readers are referred to Ref.\onlinecite{hirayama17} for ways of renormalizing the nonlocal interaction for this purpose.}

Now we reached the effective Hamiltonian (\ref{HLCGW}) in the form of Eq.(\ref{Hamiltonian0}). This offers effective Hamiltonians for the L degrees of freedom to be solved by solvers beyond the DFT and GW schemes.

\subsection{Computational Conditions}

For the crystallographic parameters, we employ the experimental results reported by Ref.~\onlinecite{Putilin} for HgBa$_2$CuO$_4$  and those reported by Ref.~\onlinecite{Jorgensen}
for La$_2$CuO$_4$. For the Hg compound we take $a=3.8782$\AA
 and $c=9.5073$\AA. The height of Ba atom measured from CuO$_2$ plane is $0.2021c$ and the apex oxygen height is $0.2940c$  The lattice constants we used for the La compounds are  $a=3.7817$\AA
 and $c=13.2487$\AA, while La and apex oxygen heights measured from the CuO$_2$ plane are $0.3607c$ and $0.1824c$, respectively. Other atomic coordinates are determined from the crystal symmetry.

Computational conditions are as follows.
The band structure calculation is based on the full-potential
LMTO implementation\cite{methfessel}. The exchange correlation
functional is obtained by the local density approximation of the
Cepeley-Alder type\cite{ceperley}) and spin-polarization is neglected.
The self-consistent LDA calculation is done for
the 12 $\times$ 12  $\times$ 12 $k$-mesh.
The muffintin (MT) radii are as follows:
$R^{\text{MT}}_{\text{Hg(HgBa2CuO4)}}=$ 2.6 bohr, $R^{\text{MT}}_{\text{Ba(HgBa2CuO4)}}=$ 3.6 bohr, 
$R^{\text{MT}}_{\text{Cu(HgBa2CuO4)}}=$ 2.15 bohr, $R^{\text{MT}}_{\text{O1(HgBa2CuO4)}}=$ 1.50 bohr (in CuO$_2$ plane), 
$R^{\text{MT}}_{\text{O2(HgBa2CuO4)}}=$ 1.10 bohr (others),
$R^{\text{MT}}_{\text{La(La2CuO4)}}=$ 2.88 bohr, $R^{\text{MT}}_{\text{Cu(La2CuO4)}}=$ 2.09 bohr,
$R^{\text{MT}}_{\text{O1(La2CuO4)}}=$ 1.40 bohr (in CuO$_2$ plane), $R^{\text{MT}}_{\text{O2(La2CuO4)}}=$ 1.60 bohr (others).
The angular momentum cutoff is taken at $l=4$ for all the sites.

The cRPA and GW calculations use a mixed basis consisting of products of two atomic orbitals and interstitial plane
waves~\cite{schilfgaarde06}.
In the cRPA and GW calculation, 
the 6 $\times$ 6 $\times$ 3 $k$-mesh is employed for the Hg compound
and the 6 $\times$ 6 $\times$ 4 $k$-mesh is employed for the La compound.
By comparing the calculations with the smaller $k$-mesh, we checked that these conditions give well converged results.
For the Hg/La compound, we include bands in [$-26.4$: $122.7$] eV (193 bands)/[$-67.6$: $126.6$] eV (134 bands)
for calculation of the screened interaction and the self-energy.
For entangled bands, we disentangle the target bands from the global KS-bands\cite{miyake09}.


\section{Result}


\subsection{HgBa$_2$CuO$_{4}$}

\begin{figure}[h]
\centering 
\includegraphics[clip,width=0.4\textwidth ]{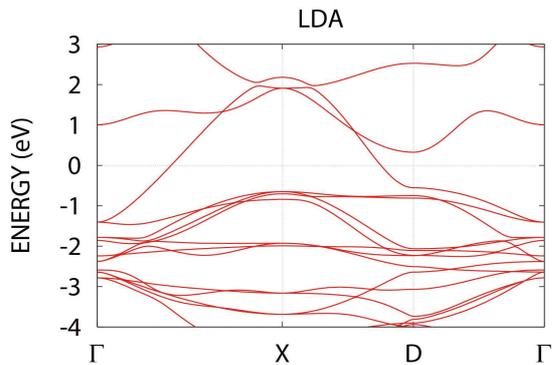} 
\caption{(Color online) Electronic band structures of HgBa$_2$CuO$_4$ obtained by the LDA.
The zero energy corresponds to the Fermi level.}
\label{bndHgLDA}
\end{figure}

\begin{figure}[h]
\centering 
\includegraphics[clip,width=0.4\textwidth ]{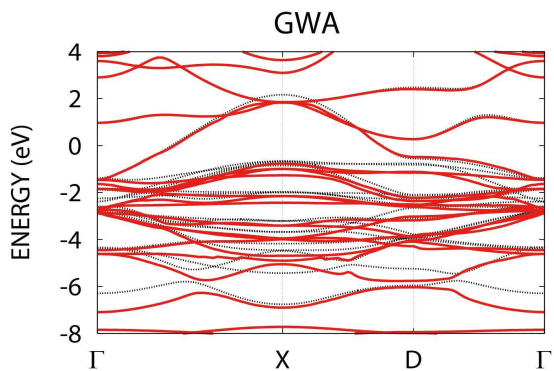} 
\caption{(Color online) 
Electronic band structures of HgBa$_2$CuO$_4$ obtained by the GWA (red solid line).
Self-energy is calculated only for the 17 bands originating from the Fe $3d$ and O $2p$ orbitals near the Fermi level.
The zero energy corresponds to the Fermi level. 
For comparison, the LDA band structure is also given (black dotted line).
}
\label{bndHgGW}
\end{figure}

Band structures  of HgBa$_2$CuO$_4$ obtained by the DFT calculations are 
shown in Fig.~\ref{bndHgLDA}.
The $17$ bands originating from the Cu $3d$ and O $2p$ orbitals exist near the Fermi level \tr{as shown in Fig.~\ref{bndHgGW}}.
The octahedral crystal field of the O atoms splits the energy of the Cu $3d$ orbital into lower $t_{2g}$ and slightly split $e_g$.
Since the electronegativity of Cu is relatively large,
the Cu $e_g$ orbitals form strong $\sigma $ covalent bonds with the O $2p$.
The bottom/top of the $17$ bands at the X point 
is the $\sigma $ bonding/anti-bonding state between the Cu $x^2-y^2$ orbital and the O $2p$ orbital.
The $s$-bands originating from Hg and Ba exist above the $17$ bands and are partially hybridized with the Cu $x^2-y^2$ anti-bonding band around the X point.

In order to improve the band structure from the LDA,
we construct the $17$ Wannier functions from the $20$ bands near the Fermi level 
(17 bands originating from the Cu $3d$  the O $2p$ orbitals and unoccupied lowest $3$ bands)
and perform the GW calculation for the $17$ bands near the Fermi level.
The Fermi level for the $17$ bands is defined by the occupation number.
Bands other than the $17$ bands are diagonalized again~\cite{miyake09}.
Since the hybridization between the $s$ band and the $17$ bands is somewhat large, 
we set the inner window for the Wannier function from the bottom of the $17$ bands to the Fermi level. 
If inner window is not set, a large Fermi surface originating from the $s$ orbitals appears. 
Due to self-energy correction of the GWA,
the difference between on-site potentials of the Cu $3d$ orbitals and the O $2p$ orbitals with different localization strengths increases
and the bandwidth of the whole $17$ band becomes larger.
Such a change in the band structure reduces the screening effect.
Moreover, each bandwidth shrinks by self energy correction.
These two effects, both the reduction of the screening effect and the shrinkage of the band width, make the correlation of the system stronger.
Below we will discuss the derivations of three types of effective Hamiltonians,
two-band effective Hamiltonian originating from the $e_g$ orbitals,
one-band effective Hamiltonian originating from the Cu $x^2-y^2$ orbital,
and three-band effective Hamiltonian originating from the Cu $x^2-y^2$ orbital and the two O $2p$ orbitals.

\tred{Recent self-consistent GW calculation\cite{Jang} indicates narrower bands than the present GW results, because of better consideration of the correlation effect, while the present study aims at much better framework by qualitatively improving the treatment of the strong correlation effect by leaving it for low-energy solvers.}   


\subsubsection{two-band Hamiltonian}

To obtain the two-band effective Hamiltonian originating from the Cu $e_g$ orbitals,
we construct the maximally localized Wanneir functions disentangled from the other $17$ bands. 
Ignoring the effect of hybridization whose energy scale is smaller than that of effective interaction of the $x^2-y^2$ anti-bonding orbital, 
we set the energy window for Wannier function as wide as possible (excluding bottom 3 bands compared to the case in the GWA for 17 bands). \tred{The three bands contain those mainly originating from the bonding and nonbonding orbitals resulted from the Cu $x^2-y^2$ and in-plane O $2p_{\sigma}$ orbitals. By excluding the three bands, we are able to construct with the correct character of the antibonding band.} 
The parameters of the main $x^2-y^2$ orbital are highly insensitive to the window width.
Effective interaction changes by less than 5 \% even when we change the number of bands in the window by two or three.
On the other hand, although the parameters of the $3z^2-r^2$ orbital change by the definition of the window,
as will be described later, the screening effect from the $3z^2-r^2$  orbital to the $x^2-y^2$ orbital is very small and \tm{the parameters for the $x^2-y^2$ orbital change only little between different choices of the windows.}
\tbb{Examples of Wannier functions of the two-band Hamiltonian is shown in Fig.~\ref{wanHg2}(a) and (b) and their spreads are listed in Supplementary Material~\footnotemark[1].}

\tr{As an alternative choice for the two-band Hamiltonian, one can exclude the bonding orbital generated from the hybridization of the  $3z^2-r^2$ and the apex oxygen $2p_z$ orbitals to constitute one of the two bands explicitly by the antibonding orbital constructed from the  Cu $3z^2-r^2$ and the apex oxygen $2p_z$ orbitals. For this choice, we exclude lowest 6 bands among 17 bands for constructing the Wannier orbitals so that the bonding orbital is excluded. \trr{This  generates substantially smaller interactions for the  $3z^2-r^2$ band. The resultant parameters are listed in Appendices \ref{AppendixA}. We show it only for the La compound because of the following reason: The two choices of the two-band Hamiltonian} may not lead to an appreciable difference in the final result because the contribution from the  $3z^2-r^2$ band is limited in the Hg compound but for the La compound, it is a subtle issue as we discuss in Sec.\ref{DiscussionA}. In principle, the final solution for the physical properties is expected to be insensitive to the two choices. }

Band structure originating from the Wannier function is shown in Fig.~\ref{bndHgGWwan2}.
Upper band around the Fermi level originates from the $x^2-y^2$ orbital,
and the lower band originates from the $3z^2-r^2$ orbital.
The $x^2- y^2$ orbital extending in the CuO$_2$ plane has a large bandwidth,
while the $3z^2-r^2$ orbital has a flat band structure.

The one-body parameters obtained as expectation values in the GWA is shown in Table~\ref{paraHg2}.
\tr{Note that the signs of the transfers for crystallographically equivalent pairs are determined from the signs of orbitals in the convention shown in Fig. \ref{tsign3}.}
The difference of the on-site potential between the $e_g$ orbitals is \tg{5.0} eV.
The position of apex oxygen varies depending on the type of the block layer.
In the Hg system, it makes the crystal field splitting of the $e_g$ orbits large.
The nearest neighbor hopping of the $x^2-y^2$ orbital is \tg{-0.43} eV, and the next-nearest neighbor hopping is \tg{0.10} eV.
Since the $x^2-y^2$ orbital extends to the (100) and (010) directions,
the third neighbor hopping is somewhat large ($-0.05$ eV).
All of the hoppings of  the $3z^2-r^2$ orbital are small.
One of the most important consequences expected from the parameters of the two-band Hamiltonian is that
the screening effect from the $3z^2-r^2$ orbital to the $x^2-y^2$ orbital would be very small.
The nearest neighbor hopping between the different $e_g$ orbitals is as small as 0.08 eV.
In addition, both on-site and next-nearest neighbor hopping are exactly 0 from the symmetry reason.
Moreover, as mentioned above, the difference in the on-site potential between the $e_g$ orbitals is not small,
so the polarization between the $e_g$ orbitals is very small.
Then the occupation number of the $3z^2-r^2$/$x^2-y^2$ orbital is nearly full/half filling, respectively.

Band structure in the cGW+SIC is shown in Fig.~\ref{bndHgGWcGWSIC2}.
Corresponding one-body parameters in the cGW+SIC are listed in Table~\ref{paraHg2}.
Since the cGW+SIC method considers only the correlation effect (self energy) of the high-energy contribution to remove the double counting of the correlation effect between the low-energy degree-of-freedom,
the one-body parameters are different from those obtained from the expected value of the Wannier orbital calculated from the full GW calculation. 
The difference of the on-site potential becomes larger than that in the Wannier 's expectation value because of the absence of the correlation within the target bands.
In addition to the increase of the on-site potential difference,
the nearest neighbor hopping between the different $e_g$ orbitals is reduced to less than half compared with that in the Wannier's expectation value,
so that the screening effect from the $3z^2-r^2$ orbital to the  $x^2-y^2$ orbital would be almost negligible in the cGW+SIC Hamiltonian.
The parameters within the same orbital do not change so appreciably. The nearest neighbor and third neighbor hoppings of the $x^2-y^2$ orbital are about the same as those calculated by the Wannier 's expectation value.
The next-nearest neighbor hopping is, however, about 40 \% larger.
The band originating from the $3z^2-r^2$ orbital is flat as is the case with the Wannier 's expectation value.
\tm{More detailed parameters beyond 10 meV are listed in Supplementary Material~\footnotemark[1]\footnotetext[1]{See Supplementary Material for more complete list of parameters including those with small values}. Longer ranged hoppings are smaller than 10meV.}

The two-body parameters are also shown in Table~\ref{paraHg2}.
The bare onsite and intraorbital Coulomb interaction of the $3z^2-r^2$/$x^2-y^2$ orbital is 24/17 eV, respectively.
The Coulomb interaction is largely screened by the bands other than the target ones,
and the energy scale is reduced by one order of magnitude.
The effective interaction of the $3z^2-r^2$/$x^2-y^2$ orbital is \tg{6.9/4.5} eV, respectively.
The effective exchange interaction is \tg{0.73} eV.
The effective interaction between adjacent sites is about 20 \% \tp{(\tg{11 \%})} of the on-site effective interaction \tg{for the $x^2-y^2$ ($3z^2-r^2$) orbital}. \tm{
More detailed longer range interactions beyond 50meV are listed in Supplementary Material~\footnotemark[1].}
The on-site effective interaction over the absolute value of the nearest neighbor hopping is about 10,
and the correlation effect of the system is very strong.
\tm{
More detailed longer range interactions beyond 50meV are listed in Supplementary Material~\footnotemark[1].}
%
\begin{figure}[h]
\begin{center}
\includegraphics[width=0.5\textwidth ]{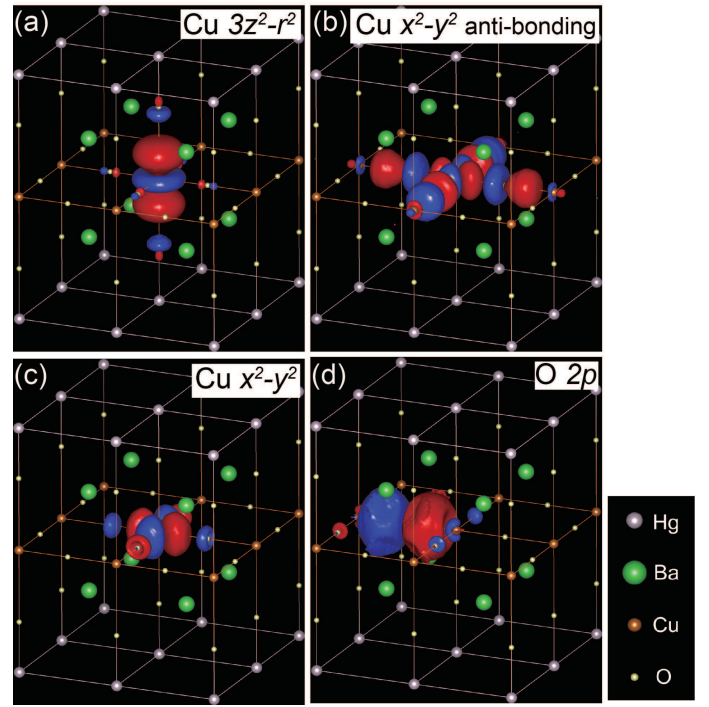} 
\caption{\tbb{(Color online) 
Isosurface of the maximally localized Wannier function for $\pm 0.03$ a.u
for (a) the Cu  $3z^2-r^2$ orbital and (b) the Cu $x^2-y^2$ anti-bonding orbital of two-band Hamiltonian
and (c) the Cu $x^2-y^2$ orbital and (d) the O $2p$ orbital of three-band Hamiltonian in HgBa$_2$CuO$_4$. The dark shaded surfaces (color in blue)
indicate the positive isosurface at +0.03 and the light shaded surfaces
(color in red) indicate -0:03.
}
}
\label{wanHg2}
\end{center}
\end{figure} 
%
%
\begin{figure}[h]
\centering 
\includegraphics[clip,width=0.4\textwidth ]{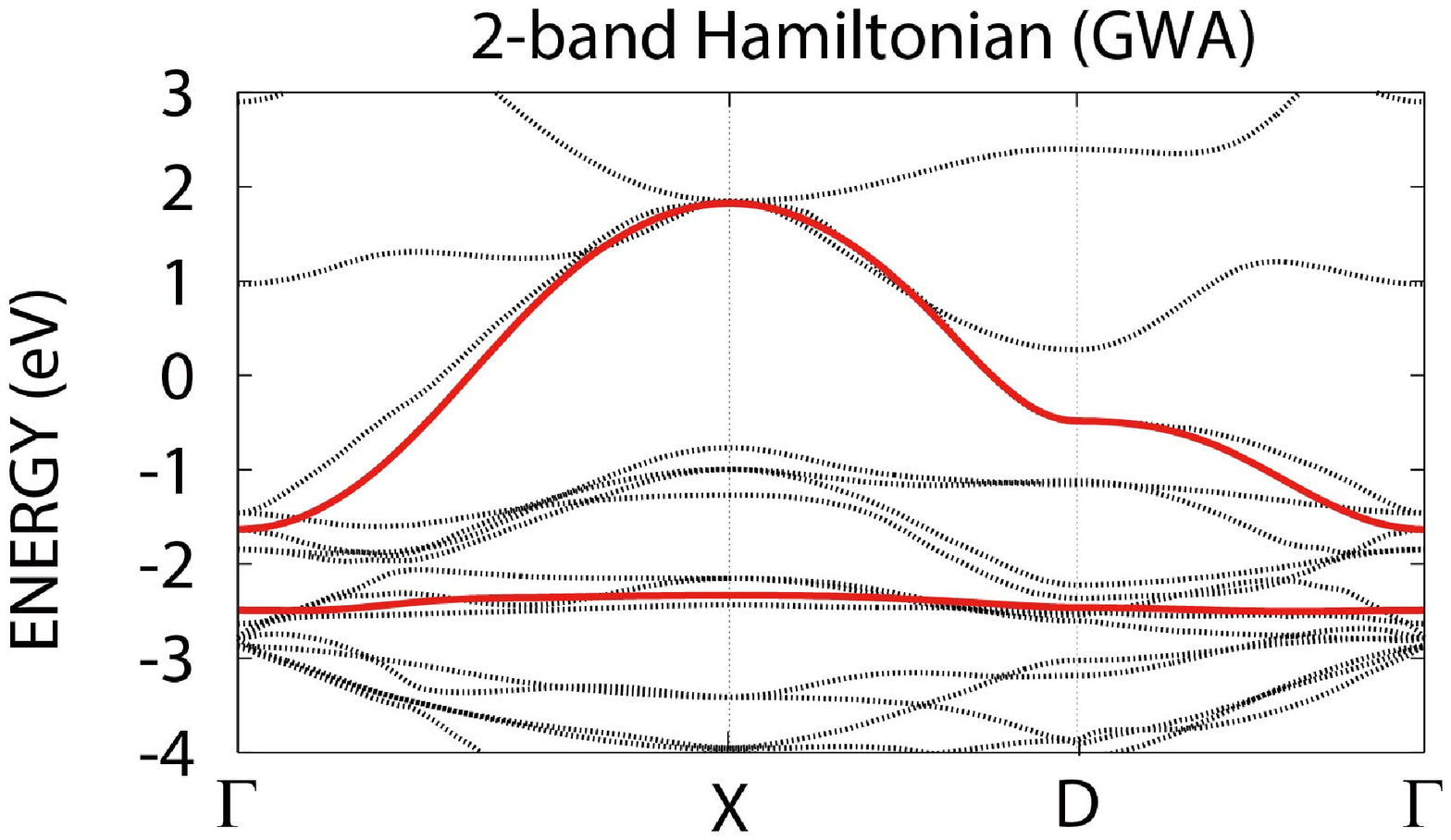} 
\caption{(Color online) Electronic band structure of two-band Hamiltonian in the GWA originating from the Cu $e_g$ Wannier orbitals for HgBa$_2$CuO$_4$.
The zero energy corresponds to the Fermi level. 
For comparison, the $17$ band structures near the Fermi level in the GWA is also given (black dotted line).
}
\label{bndHgGWwan2}
\end{figure} 
%

\begin{table*}[h] 
\caption{
Transfer integral and effective interaction in two-band Hamiltonian for HgBa$_2$CuO$_4$ (in eV).
We show the transfer integral in the GWA as well as in the cGW-SIC for comparison, while the effective interaction is same in both the GWA and the cGW-SIC.
$v$ and $J_{v}$ represent the bare Coulomb interaction/exchange interactions respectively. $U(0)$ and $J(0)$ represent the static values of the effective Coulomb interaction/exchange interactions (at $\omega=0$). The index 'n' and 'nn' represent the nearest unit cell [1,0,0] and the next-nearest unit cell [1,1,0] respectively. 
The occupation number in the GWA is also given in this Table.
}
\ 
\label{paraHg2} 
\begin{tabular}{c|cc|cc|cc|cc|cc} 
\hline \hline \\ [-8pt]
$t $(GWA)   &       &  $(0,0,0)$  &      & $(1,0,0)$ &       & $(1,1,0)$ &      &  $(2,0,0)$     \\ [+1pt]
\hline \\ [-8pt]
      &  $3z^2-r^2 $ &  $x^2-y^2 $  &  $3z^2-r^2 $ &  $x^2-y^2 $ &  $3z^2-r^2 $ &  $x^2-y^2 $ &  $3z^2-r^2 $ &  $x^2-y^2 $ \\ 
\hline \\ [-8pt] 
$3z^2-r^2 $  &  -2.282 & 0.000 &  -0.018 & 0.084 &  -0.006 & 0.000 & -0.003 &  0.010 \\
$x^2-y^2 $  &  0.000  & 0.144 & 0.084  &-0.453 & 0.000   &  0.074 &  0.010 & -0.051 \\
\hline \hline \\ [-8pt]
$t $(cGW-SIC)   &       &  $(0,0,0)$  &      & $(1,0,0)$ &        & $(1,1,0)$ &        &  $(2,0,0)$     \\ [+1pt]
\hline \\ [-8pt]
      &  $3z^2-r^2 $ &  $x^2-y^2 $  &  $3z^2-r^2 $ &  $x^2-y^2 $ &  $3z^2-r^2 $ &  $x^2-y^2 $ &  $3z^2-r^2 $ &  $x^2-y^2 $ \\ 
\hline \\ [-8pt] 
$3z^2-r^2 $  & -3.811 & 0.000 & 0.013 & 0.033  & -0.003 & 0.000 & 0.000 & 0.002 \\
$x^2-y^2 $  & 0.000  & 0.197 & 0.033 & -0.426 & 0.000  & 0.102 & 0.002 & -0.048 \\
\hline \hline \\ [-8pt]  
   &       &  $v$  &      & $U(0)$ &     & $J_{v}$ &       &  $J(0)$      \\ [+1pt]
\hline \\ [-8pt]
       &  $3z^2-r^2 $ &  $x^2-y^2 $  &  $3z^2-r^2 $ &  $x^2-y^2 $ &  $3z^2-r^2 $ &  $x^2-y^2 $ &  $3z^2-r^2 $ &  $x^2-y^2 $ \\ 
\hline \\ [-8pt] 
$3z^2-r^2 $   & 24.348  & 18.672 &  6.922  & 3.998 &             &   0.808   &               &  0.726 \\
$x^2-y^2 $  & 18.672  & 17.421 & 3.998   & 4.508 & 0.808  &                &  0.726   &             \\ 
\hline \hline \\ [-8pt]  
       &       &  $v_{\text{n}}$ &    & $V_{\text{n}}(0)$ &       & $v_{\text{nn}}$  &       &  $V_{\text{nn}}(0)$     \\ [+1pt]
\hline \\ [-8pt] 
      &  $3z^2-r^2 $ &  $x^2-y^2 $  &  $3z^2-r^2 $ &  $x^2-y^2 $ &  $3z^2-r^2 $ &  $x^2-y^2 $ &  $3z^2-r^2 $ &  $x^2-y^2 $ \\ 
\hline \\ [-8pt] 
$3z^2-r^2 $  & 3.669 &  3.922  & 0.764   & 0.833 & 2.657  & 2.696 & 0.486 &  0.502 \\
$x^2-y^2 $  &   3.922 &  4.155 & 0.833  & 0.901&  2.696 & 2.749 & 0.502 & 0.522 \\
\hline \\ [-8pt]
occ.(GWA)      &  $3z^2-r^2 $ &  $x^2-y^2 $  \\ 
\hline \\ [-8pt] 
                     & 1.992  &  1.008   \\
\hline
\hline 
\end{tabular} 
\end{table*} 
%
\begin{figure}[h]
\centering 
\includegraphics[clip,width=0.4\textwidth ]{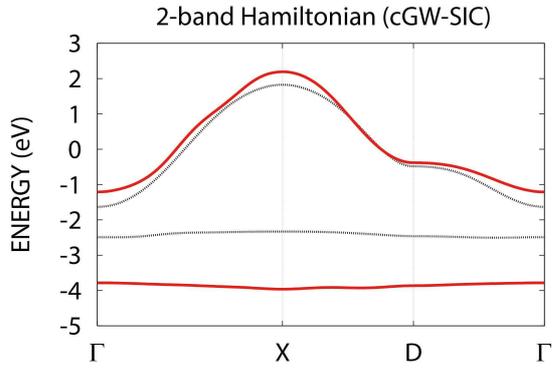} 
\caption{(Color online) Electronic band structure of two-band Hamiltonian in the cGW-SIC originating from the Cu $e_g$ Wannier orbitals for HgBa$_2$CuO$_4$.
The zero energy corresponds to the Fermi level. 
For comparison, the band structure in the GWA is also given (black dotted line).
}
\label{bndHgGWcGWSIC2}
\end{figure} 

\subsubsection{one-band Hamiltonian}

\begin{figure}[h]
\centering 
\includegraphics[clip,width=0.4\textwidth ]{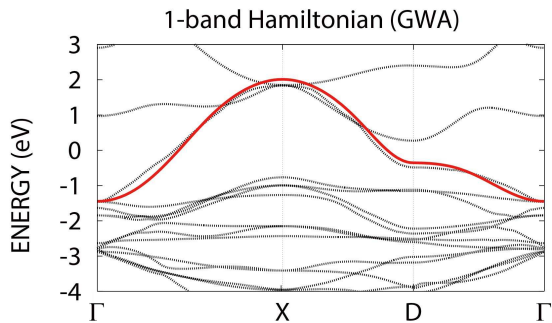} 
\caption{(Color online) Electronic band structure of one-band Hamiltonian in the GWA originating from the Cu $d_{x^2-y^2}$ Wannier orbital for HgBa$_2$CuO$_4$.
The zero energy corresponds to the Fermi level. 
For comparison, the $17$ band structures near the Fermi level in the GWA is also given (black dotted line).
}
\label{bndHgGWwan1}
\end{figure}

We use the same Wannier function of the $x^2-y^2$ orbital in the one-band Hamiltonian as that in the two-band Hamiltonian.
\tred{This is because the largest energy window for the construction of the maximally localized Wannier orbital by keeping the physically correct antibonding orbital for the $x^2-y^2$ orbital is the same as the two-band construction (the 14-band window).
}
Unlike the two-band Hamiltonian,
since only the $x^2-y^2$ orbital is disentangled from the entire band,
the hybridization between the $3z^2-r^2$ orbital and other orbitals except the $x^2 - y^2$ orbital is retained.
Band structure originating from the Wannier function of the $x^2-y^2$ orbital is shown in Fig.~\ref{bndHgGWwan1}.
This is exactly the same as that of the two-band Hamiltonian.
Corresponding one-body parameters are listed in the upper row of Table~\ref{paraHg1}.

\begin{figure}[h]
\centering 
\includegraphics[clip,width=0.4\textwidth ]{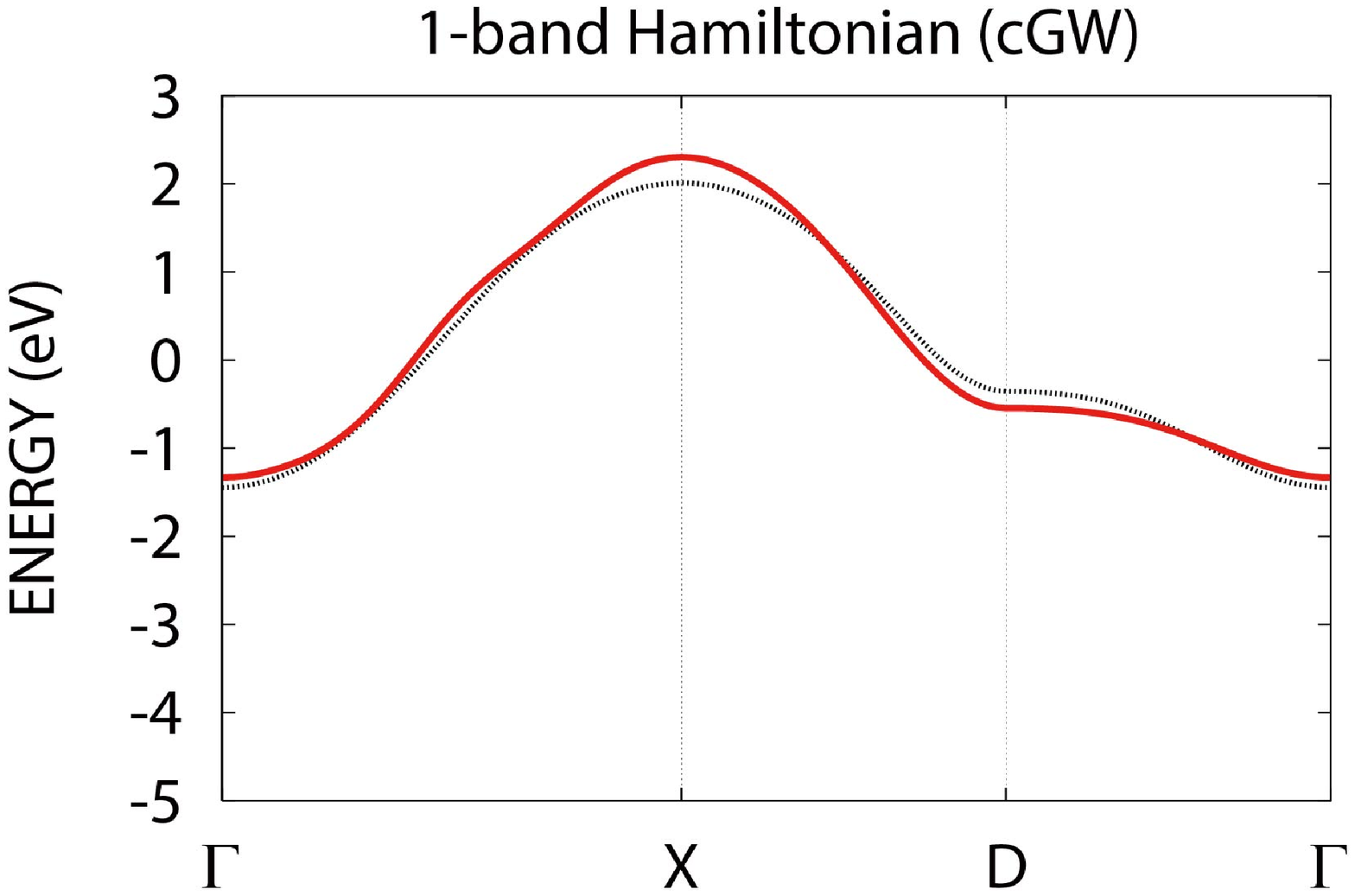} 
\caption{(Color online) Electronic band structure of one-band Hamiltonian in the cGW originating from the Cu $d_{x2-y^2}$ Wannier orbital for HgBa$_2$CuO$_4$.
The zero energy corresponds to the Fermi level. 
For comparison, the band structure in the GWA is also given (black dotted line).
}
\label{bndHgGWcGW1}
\end{figure}

Band structure in the cGW is shown in Fig.~\ref{bndHgGWcGW1}.
In the case of the single band Hamiltonian, there is no need to consider SIC.
The one-body parameter in the cGW+SIC and the two-body parameter obtained from the cRPA are listed in the second row group of Table~\ref{paraHg1}. \tr{Parameters for longer ranged pairs up to the unit cell distance $(3,3,0)$ are given in Supplementary Material~\footnotemark[1]. Beyond $(3,3,0)$, one-body parameters are all below 10 meV, and the two-body parts beyond $(3,3,0)$ can be estimated from the $1/r$ dependence both for Hg and La compounds.}
\tm{The difference from the one-body parameters of the $x^2-y^2$ orbital for the two-band Hamiltonian is small.
This is because the polarization effect from the $3z^2-r^2$ orbital to the $x^2-y^2$ orbital is significantly small from both the symmetry and energy reasons,
as is addressed in the above analyses of the two-band Hamiltonian.}

%
\begin{table*}[ptb] 
\caption{
Transfer integral and effective interaction of one-band Hamiltonian for HgBa$_2$CuO$_4$ (in eV).
We show the transfer integrals in the GWA as well as in the cGW for comparison, while the effective interactions are the same in both the GWA and the cGW.
$v$ represents the bare Coulomb interaction.
$U(0)$represent the static values of the effective Coulomb interaction (at $\omega=0$).
The index 'n' and 'nn' represent the nearest unit cell [1,0,0] and the next-nearest unit cell [1,1,0] respectively. 
}
\ 
\label{paraHg1} 
\begin{tabular}{c|c|c|c|c|c|c|c} 
\hline \hline \\ [-8pt]
$t $(GWA)   &     $(0,0,0)$  &   $(1,0,0)$ &     $(1,1,0)$ &   $(2,0,0)$     \\ [+1pt]
\hline \\ [-8pt] 
  $x^2-y^2$    &   0.164 & -0.453 &0.074 &  -0.051  \\
\hline \hline \\ [-8pt]
$t $(cGW)   &      $(0,0,0)$  &   $(1,0,0)$ &     $(1,1,0)$ &      $(2,0,0)$    \\ [+1pt]
\hline \\ [-8pt] 
$x^2-y^2$  & 0.190 &-0.461 &  0.119 & -0.072 \\
\hline \hline \\ [-8pt]  
               &      $v$  &     $U(0)$ &     $v_{\text{n}}$ &    $V_{\text{n}}(0)$ &   $v_{\text{nn}}$  &     $V_{\text{nn}}(0)$    \\ [+1pt]
\hline \\ [-8pt]
$x^2-y^2$   &  17.421  & 4.374 &  4.155  & 1.093 &2.749  & 0.736 \\
\hline
\hline 
\end{tabular} 
\end{table*} 

\subsubsection{three-band Hamiltonian}

The three-band Hamiltonian consists of the Cu $3d$ and O $2p$ orbitals.
We set the energy window for the maximally localized Wannier functions as same as that in the previous calculation of the GWA.
\tbb{The Wannier functions of the three-band Hamiltonian are illustrated in Fig.~\ref{wanHg2}(c) and (d) and their spreads are listed in Supplementary Material~\footnotemark[1].}
The three Wannier orbitals are close to the Cu $x^2-y^2$ and O $2p$ atomic orbitals.

\begin{figure}[h]
\centering 
\includegraphics[clip,width=0.4\textwidth ]{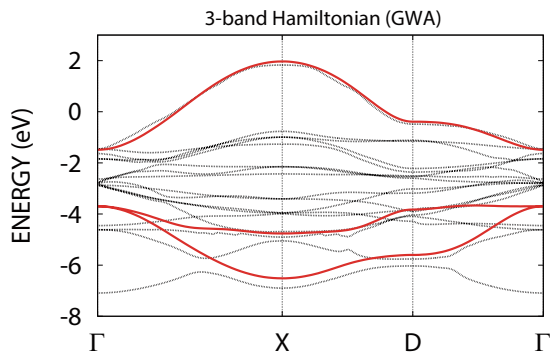} 
\caption{(Color online) Electronic band structure of three-band Hamiltonian in the GWA originating from the Cu $d_{x^2-y^2}$ and O $2p$ Wannier orbitals for HgBa$_2$CuO$_4$.
The zero energy corresponds to the Fermi level. 
For comparison, the $17$ band structures near the Fermi level in the GWA is also given (black dotted line).
}
\label{bndHgGWwan3}
\end{figure}

\begin{figure}[h]
\centering 
\includegraphics[clip,width=0.4\textwidth ]{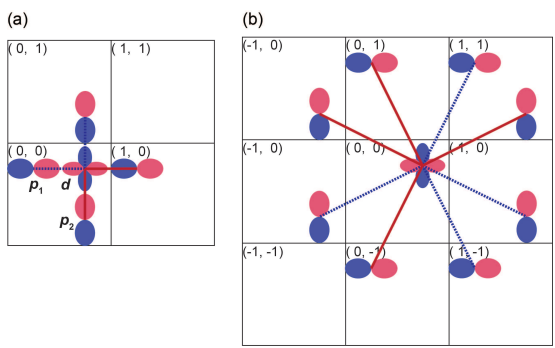} 
\caption{(Color online) Sign of the transfer integral between the Cu $d_{x^2-y^2}$ and O $2p$ orbitals  for three-band Hamiltonian
for (a) the nearest-neighbor hopping and (b) the next-nearest-neighbor hopping. \tg{Red and blue colors show opposite signs of the wavefunction.}
}
\label{tsign3}
\end{figure}

Band structure calculated from the Wannier functions is shown in Fig.~\ref{bndHgGWwan3}.
Although the Wannier functions are close to the atomic orbitals,
in the three-band Hamiltonian, bonding, nonbonding and anti-bonding bands generated from the Cu $x^2-y^2$ and the O $2p$ orbitals are naturally formed because of the strong hybridization between the $d$ and $p$ orbitals.
The highest band closest to the Fermi level in the GWA consists of the anti-bonding orbital constructed from the Cu $x^2-y^2$ and the O $2p_{\sigma}$ orbitals.
The lower two bands are the O $2p$ non-bonding and bonding bands.
At the $\Gamma $ point, due to the symmetry, hybridization between the three orbitals completely disappears
and the O $2p$ band degenerates.

Corresponding one-body parameters of the Wannier function are listed in the upper rows of Table~\ref{paraHg3}.
The difference in the on-site potentials between the Cu $x^2-y^2$ and O $2p$ orbitals  is \tg{2.4} eV.
The nearest neighbor hopping between the Cu $x^2-y^2$ and O $2p$ orbitals reaches \tg{1.26} eV, making a large splitting of bonding and anti-bonding bands.
The nearest neighbor hopping between the two nearest O $2p$ orbitals is also large, \tg{$0.75$} eV.
Long range hopping in the two and one-band Hamiltonians has a relatively large amplitudes through the hybridization with the O $2p$ orbitals.
In contrast, in the three-band Hamiltonians, the direct long range hopping between the atomic orbital-like Cu $x^2-y^2$ orbital is relatively small.
The occupation number of the Cu $x^2-y^2$/O $2p$ orbital is $\sim$ 1.4/1.8, respectively.
The deviation from the full filling of the occupancy number of the O $2p$ orbital arises from the hybridization.

\begin{figure}[h]
\centering 
\includegraphics[clip,width=0.4\textwidth ]{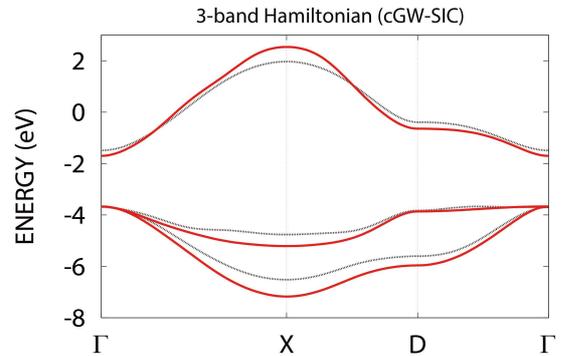} 
\caption{(Color online) Electronic band structure of three-band Hamiltonian in the cGW-SIC originating from the Cu $d_{x^2-y^2}$ and O $2p$ Wannier orbitals for HgBa$_2$CuO$_4$.
The zero energy corresponds to the Fermi level. 
For comparison, the band structure in the GWA is also given (black dotted line).
}
\label{bndHgGWcGWSIC3}
\end{figure}

Band structure in the cGW+SIC is shown in Fig.~\ref{bndHgGWcGWSIC3}.
Corresponding one-body parameters in the cGW+SIC are listed in the second group of rows of Table~\ref{paraHg3}.
The difference in the on-site potential between the Cu $x^2-y^2$ and O $2p$ orbitals (2.4 eV) is nearly the same as that in the GWA.
The nearest neighbor hopping between the Cu $x^2-y^2$ and O $2p$ orbitals with the energy scale of 1 eV exhibits several \% ($\sim$ 70 meV) increase from the GWA result  
and the nearest neighbor hopping between the O $2p$ orbitals also increases by 100 meV compared to that in the GWA,
which make the energy splitting between the bonding and anti-bonding states at the X point larger than that in the GWA.
Longer range hoppings in the cGW+SIC with the energy scale of 10 meV also increase compared to those in the GWA.
\tm{Further neighbor hoppings larger than 10meV are listed in Supplementary Material~\footnotemark[1].}
The two-body parameters are also listed in Table~\ref{paraHg3}.
Effective on-site interaction of both the Cu $x^2-y^2$ and O $2p$ orbitals is reduced to about 30 \% of the bare on-site interactions.
The nearest neighbor effective interaction between the Cu $x^2-y^2$ and O $2p$ orbitals is large, about 2 eV.
The other interactions are 1 eV or less, \tm{and the further neighbor interactions beyond the next nearest neighbors gradually decrease with approximately $1/r$ behavior and are listed in the Supplementary Material~\footnotemark[1].}

\begin{table*}[h] 
\caption{
Transfer integrals and effective interactions for three-band Hamiltonian of HgBa$_2$CuO$_4$ (in eV).
We show the transfer integral in the GWA as well as in the cGW-SIC for comparison, while the effective interaction is the same in both the GWA and the cGW-SIC.
$v$ and $J_{v}$ represent the bare Coulomb and exchange interactions, respectively. $U(0)$ and $J(0)$ represent the static values of the effective Coulomb and exchange interactions, respectively (at $\omega=0$). The index 'n' and 'nn' represent the nearest, [1,0,0] and the next-nearest sites [1,1,0] respectively. 
The occupation number in the GWA is also given in the bottom column ``occu.(GWA)'' in this Table.
}
\ 
\label{paraHg3} 
\begin{tabular}{c|ccc|ccc|ccc|ccc|ccc} 
\hline \hline \\ [-8pt]
$t $(GWA)   &       &  $(0,0,0)$  &       &     & $(1,0,0)$ &    &       & $(1,1,0)$ &      &     &  $(2,0,0)$ &     \\ [+1pt]
\hline \\ [-8pt]
      &  $x^2-y^2$ &  $p_1$ &  $p_2$ & $x^2-y^2$ &  $p_1$ &  $p_2$ &  $x^2-y^2$ &  $p_1$ &  $p_2$ & $x^2-y^2$ &  $p_1$ &  $p_2$ \\ 
\hline \\ [-8pt] 
$x^2-y^2$  & -1.597 & -1.184 &  1.184   & -0.014 & -0.026 & -0.016 &  0.020 & 0.004 & -0.004 & 0.002 & -0.005 & -0.002 \\ 
$p_1$          & -1.184 & -3.909 & -0.659  &   1.184 &  0.111 &  0.659  & -0.016 & 0.039 &  0.003 & 0.026 & -0.008 &  0.003 \\
$p_2$          &  1.184 & -0.659 &  -3.909  &  -0.016 &-0.003 & -0.061 &  0.016 & 0.003 &  0.039  & -0.002 & 0.006 & -0.004    \\
\hline \hline \\ [-8pt]
$t $(cGW-SIC)   &       &  $(0,0,0)$  &       &     & $(1,0,0)$ &    &       & $(1,1,0)$ &      &     &  $(2,0,0)$ &     \\ [+1pt]
\hline \\ [-8pt]
      &  $x^2-y^2$ &  $p_1$ &  $p_2$ & $x^2-y^2$ &  $p_1$ &  $p_2$ &  $x^2-y^2$ &  $p_1$ &  $p_2$ & $x^2-y^2$ &  $p_1$ &  $p_2$ \\ 
\hline \\ [-8pt] 
$x^2-y^2$  &  -1.696 & -1.257 & 1.257  & -0.012 & -0.033 & -0.056 &  0.021 & -0.012  & 0.012 & -0.012 &  0.004 & -0.003 \\
$p_1$          & -1.257 & -4.112 & -0.751 & 1.257  &  0.181  & 0.751  & -0.056 &  0.054  & 0.004 &  0.033 & -0.006 &  0.004 \\
$p_2$          &  1.257 & -0.751 & -4.112 & -0.056 & -0.004 & -0.060  & 0.056 &  0.004  & 0.054 & -0.003 &  0.001 & -0.004 \\
\hline \hline \\ [-8pt]  
   &       &  $v$  &       &     & $U(0)$ &    &       & $J_{v}$ &      &     &  $J(0)$ &     \\ [+1pt]
\hline \\ [-8pt]
      &  $x^2-y^2$ &  $p_1$ &  $p_2$ & $x^2-y^2$ &  $p_1$ &  $p_2$ &  $x^2-y^2$ &  $p_1$ &  $p_2$ & $x^2-y^2$ &  $p_1$ &  $p_2$ \\ 
\hline \\ [-8pt] 
$x^2-y^2$ & 28.821  &  8.010 &   8.010   & 8.837 &  1.985 &   1.985  &            &  0.063 &   0.063  &           &   0.048 &  0.048  \\ 
$p_1$         &   8.010  & 17.114 &  5.319  &  1.985  & 5.311  &  1.210  &  0.063  &            &  0.041  & 0.048  &  -          &  0.020 \\
 $p_2$        &   8.010  &  5.319  & 17.114  & 1.985  &  1.210 &   5.311 &  0.063   & 0.041  &            &  0.048  & 0.020  &         \\
\hline \hline \\ [-8pt]  
       &       &  $v_{\text{n}}$ &    &     & $V_{\text{n}}(0)$ &    &       & $v_{\text{nn}}$  &      &     &  $V_{\text{nn}}(0)$ &     \\ [+1pt]
\hline \\ [-8pt] 
      &  $x^2-y^2$ &  $p_1$ &  $p_2$ & $x^2-y^2$ &  $p_1$ &  $p_2$ &  $x^2-y^2$ &  $p_1$ &  $p_2$ & $x^2-y^2$ &  $p_1$ &  $p_2$ \\ 
\hline \\ [-8pt] 
$x^2-y^2$ &   3.798 &  8.010  &  3.339  & 0.804  &  1.985   & 0.650   &  2.706  &  3.339  &  3.339  &  0.380  &  0.545  &  0.544  \\
$p_1$         &   2.577  & 3.877  &  2.417  &  0.499  &  0.847  &  0.450  &  2.172  &  2.678  &  2.417  &  0.286  &  0.415  &  0.356 \\
$p_2$        &   3.339  & 5.319  &  3.601   &  0.650  &  1.210  &  0.705  &  2.172  &  2.417  &  2.678  &  0.286  &  0.356  &  0.414 \\
\hline \\ [-8pt]
occ.(GWA)      &  $x^2-y^2$ &  $p_1$ &  $p_2$  \\ 
\hline \\ [-8pt] 
                     & 1.437 & 1.781  & 1.781   \\
\hline
\hline 
\end{tabular} 
\end{table*} 

%

\subsection{La$_2$CuO$_{4}$}

\begin{figure}[h!]
\centering 
\includegraphics[clip,width=0.4\textwidth ]{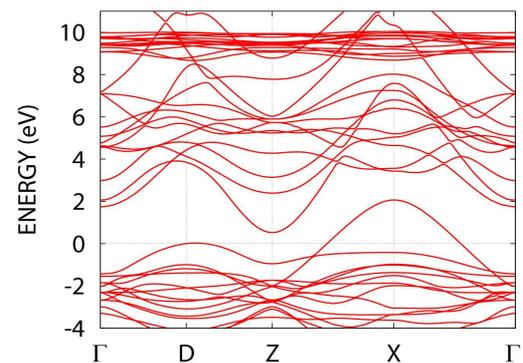} 
\caption{(Color online) Electronic band structures of La$_2$CuO$_4$ 
as a starting point of calculation, where the $4f$ band is raised up by the GW self-energy \tg{after the LDA calculation}.
The zero energy corresponds to the Fermi level.}
\label{bndLaLDA}
\end{figure}

Band structures  of La$_2$CuO$_4$ obtained by the DFT calculations are 
shown in Figs.~\ref{bndLaLDA}.
\tr{The basic framework for the derivation is the same as the La compound and we do not repeat it here.}
\begin{figure}[h!]
\centering 
\includegraphics[clip,width=0.4\textwidth ]{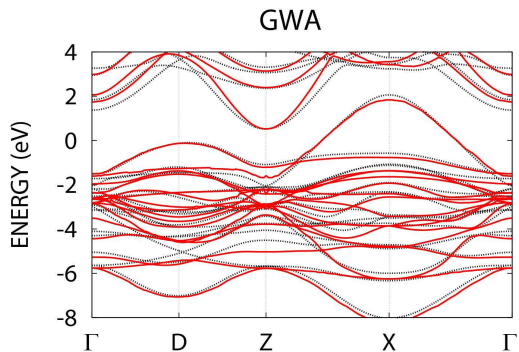} 
\caption{(Color online) Electronic band structures of La$_2$CuO$_4$ obtained by the GWA
for the $dp$ 17 bands.
The zero energy corresponds to the Fermi level. 
For comparison, 
the 0th shot band structure shown in Fig.~\ref{bndLaLDA} is also given (black dotted line).
}
\label{bndLaGW}
\end{figure} 

\subsubsection{two-band Hamiltonian}
\tr{For the two-band Hamiltonian, 
\tbb{the Wannier functions are illustrated in Fig.\ref{wanLa2}(a),(b) and their spreads are listed in Supplementary Material~\footnotemark[1].}
The band structure obtained from the full GWA is illustrated in Fig.~\ref{bndLaGWwan2}, while the cGW-SIC results are shown in Fig.~\ref{bndLaGWcGWSIC2}.  The choice of the window to construct the Wannier orbital is more subtle than the case of the Hg compound, because the $3d_{3z^2-r^2}$ orbital may play more active role. Although the window should be taken as large as possible to make the Wannier orbital maximally localized, the  ``$3d_{3z^2-r^2}$ band" may not become the hybridized antibonding band. Here we show the two-band Hamiltonian parameters derived from the Wannier orbital excluding the apex oxygen $2p_z$ atomic orbital in the main text. Another choice where  one of the Wannier orbitals is constructed from the $2p_z-3d_{3z^2-r^2}$ antibonding band is discussed in Appendix.}
%
\begin{figure}[h!]
\centering 
\includegraphics[clip,width=0.5\textwidth ]{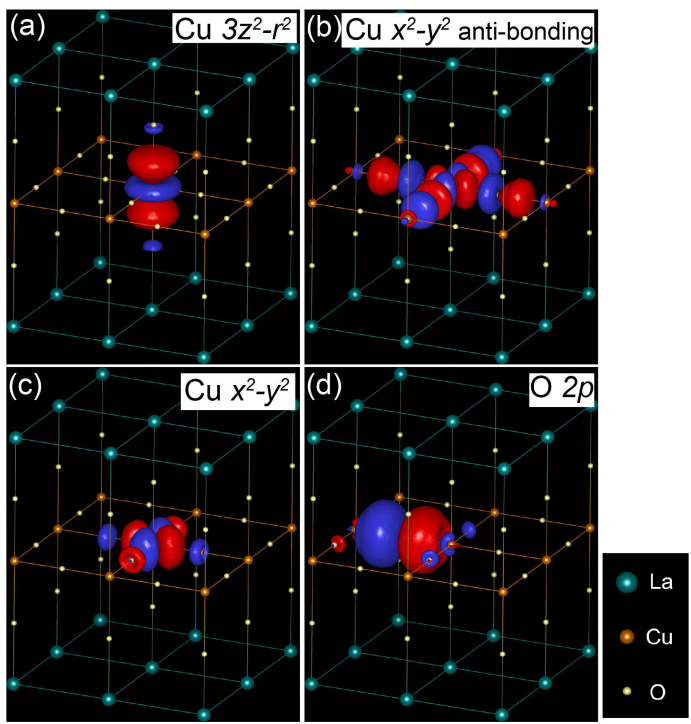} 
\caption{
\tbb{(Color online) Isosurface of the maximally localized Wannier function for $\pm 0.03$ a.u
for (a) the Cu  $3z^2-r^2$ orbital and (b) the Cu $x^2-y^2$ anti-bonding orbital of two-band Hamiltonian
and (c) the Cu $x^2-y^2$ orbital and (d) the O $2p$ orbital of three-band Hamiltonian in La$_2$CuO$_4$. Notations are the same as Fig.~\ref{wanHg2}}
}
\label{wanLa2}
\end{figure} 
%

%
\begin{figure}[h!]
\centering 
\includegraphics[clip,width=0.4\textwidth ]{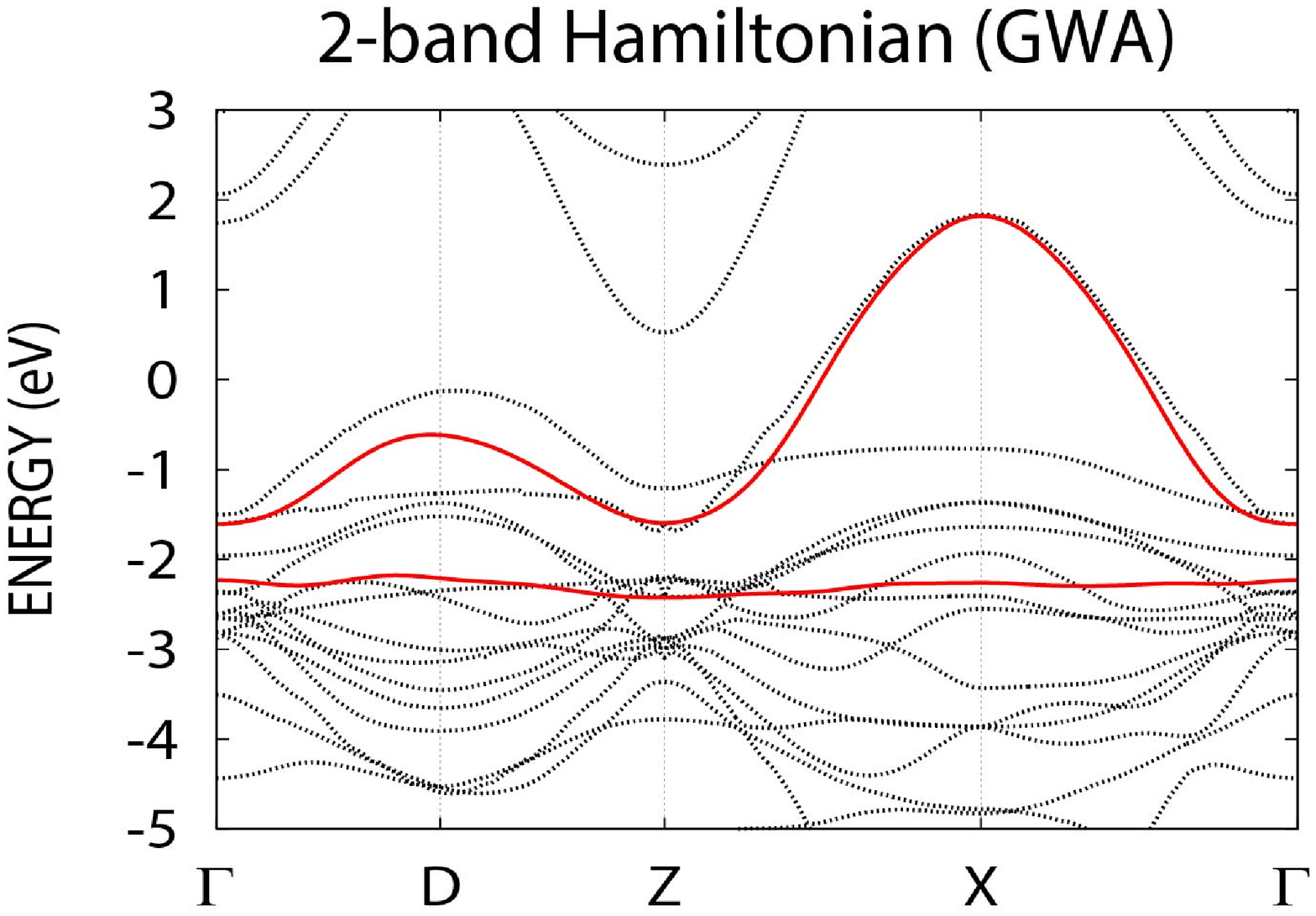} 
\caption{(Color online) Electronic band structure of two-band Hamiltonian in the GWA originating from the Cu $e_g$ Wannier orbitals for La$_2$CuO$_4$.
The zero energy corresponds to the Fermi level. 
For comparison, the $17$ band structures near the Fermi level in the GWA is also given (black dotted line).
}
\label{bndLaGWwan2}
\end{figure} 
\begin{figure}[h!]
\centering
\includegraphics[clip,width=0.4\textwidth ]{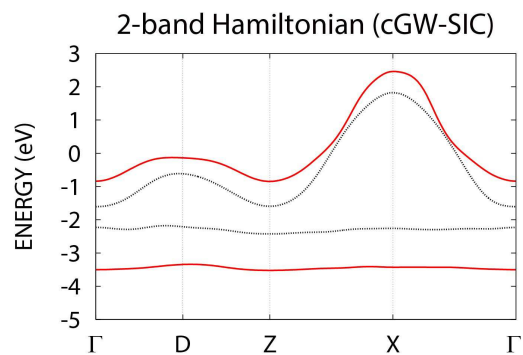} 
\caption{(Color online) Electronic band structure of two-band Hamiltonian in the cGW-SIC originating from the Cu $e_g$ Wannier orbitals for La$_2$CuO$_4$.
The zero energy corresponds to the Fermi level. 
For comparison, the band structure in the GWA is also given (black dotted line).
}
\label{bndLaGWcGWSIC2}
\end{figure} 
\tr{The obtained parameters for the two-band Hamiltonian is listed in Table~\ref{paraLa2}.  Here we show the results obtained from the choice of 14 bands by excluding 3 bands among the 17 bands for the window to determine the Wannier orbital. This means that the Wannier orbital for the antibonding band constructed from the Cu $3d_{x^2-y^2}$ and inplane oxygen $2p_{\sigma}$ band is employed, while Cu $3d_{3z^2-r^2}$ band in the two-band Hamiltonian is constructed by excluding the apex oxygen $2p_z$ orbital, because the $2p_z$ orbital constitutes another Wannier orbital orthogonal to the Cu $3d_{3z^2-r^2}$ Wannier orbital. In Appendix, we list the parameters obtained from the two-band Hamiltonian, in which one band is explicitly constructed from the antibonding $3d_{3z^2-r^2}$ and the apex oxygen $2p_z$ orbitals. This is obtained by excluding lowest 7 bands among the 17 bands for the construction window of the Wannier orbitals. The effective Hamiltonian parameters 
up to the relative unit-cell coordinate $(3,3,0)$ are listed in the Supplementary Material~\footnotemark[1] in the same way as the Hg compound.}
\begin{table*}[h] 
\caption{
Transfer integral and effective interaction in two-band Hamiltonian for La$_2$CuO$_4$ (in eV).
Notations are the same as Table~\ref{paraHg2}.
}
\ 
\label{paraLa2} 
\begin{tabular}{c|cc|cc|cc|cc|cc} 
\hline \hline \\ [-8pt]
$t $(GWA)   &       &  $(0,0,0)$  &      & $(1,0,0)$ &       & $(1,1,0)$ &      &  $(2,0,0)$     \\ [+1pt]
\hline \\ [-8pt]
      &  $3z^2-r^2 $ &  $x^2-y^2 $  &  $3z^2-r^2 $ &  $x^2-y^2 $ &  $3z^2-r^2 $ &  $x^2-y^2 $ &  $3z^2-r^2 $ &  $x^2-y^2 $ \\ 
\hline \\ [-8pt] 
$3z^2-r^2 $  & -1.996  & 0.000 &  -0.007  & 0.082 &   -0.019 &  0.000 & 0.012 & -0.002 \\
$x^2-y^2 $  &  0.000  &  0.159  &  0.082 & -0.451 & 0.000  & 0.088 &  -0.002 & -0.041 \\
\hline \hline \\ [-8pt]
$t $(cGW-SIC)   &       &  $(0,0,0)$  &      & $(1,0,0)$ &        & $(1,1,0)$ &        &  $(2,0,0)$     \\ [+1pt]
\hline \\ [-8pt]
      &  $3z^2-r^2 $ &  $x^2-y^2 $  &  $3z^2-r^2 $ &  $x^2-y^2 $ &  $3z^2-r^2 $ &  $x^2-y^2 $ &  $3z^2-r^2 $ &  $x^2-y^2 $ \\ 
\hline \\ [-8pt] 
$3z^2-r^2 $  &  -3.426 & 0.000 &  -0.008 & 0.057 &  -0.013 &  0.000 & 0.006 & 0.009 \\
$x^2-y^2 $   &  0.000  & 0.313 &  0.057 & -0.389 & 0.000 &  0.136 &  0.009 & 0.003 \\
\hline \hline \\ [-8pt]  
   &       &  $v$  &      & $U(0)$ &     & $J_{v}$ &       &  $J(0)$      \\ [+1pt]
\hline \\ [-8pt]
       &  $3z^2-r^2 $ &  $x^2-y^2 $  &  $3z^2-r^2 $ &  $x^2-y^2 $ &  $3z^2-r^2 $ &  $x^2-y^2 $ &  $3z^2-r^2 $ &  $x^2-y^2 $ \\ 
\hline \\ [-8pt] 
$3z^2-r^2 $   & 26.091 & 20.037 &  7.993  & 4.906 &             &   0.874   &               &  0.793 \\
$x^2-y^2 $  & 20.037 & 18.694 & 4.906  & 5.482 & 0.874  &                &  0.793   &             \\ 
\hline \hline \\ [-8pt]  
       &       &  $v_{\text{n}}$ &    & $V_{\text{n}}(0)$ &       & $v_{\text{nn}}$  &       &  $V_{\text{nn}}(0)$     \\ [+1pt]
\hline \\ [-8pt] 
      &  $3z^2-r^2 $ &  $x^2-y^2 $  &  $3z^2-r^2 $ &  $x^2-y^2 $ &  $3z^2-r^2 $ &  $x^2-y^2 $ &  $3z^2-r^2 $ &  $x^2-y^2 $ \\ 
\hline \\ [-8pt] 
$3z^2-r^2 $  & 3.793  & 4.021  & 1.431  & 1.497 & 2.745  & 2.779 & 1.186 &   1.196 \\
$x^2-y^2 $  &   4.021 &  4.230 & 1.497  & 1.562 &  2.779 &  2.824 & 1.196   & 1.210 \\
\hline \\ [-8pt]
occ.(GWA)      &  $3z^2-r^2 $ &  $x^2-y^2 $  \\ 
\hline \\ [-8pt] 
                     & 1.989  &  1.011   \\
\hline
\hline 
\end{tabular} 
\end{table*} 

\subsubsection{one-band Hamiltonian}
\begin{figure}[h!]
\centering 
\includegraphics[clip,width=0.4\textwidth ]{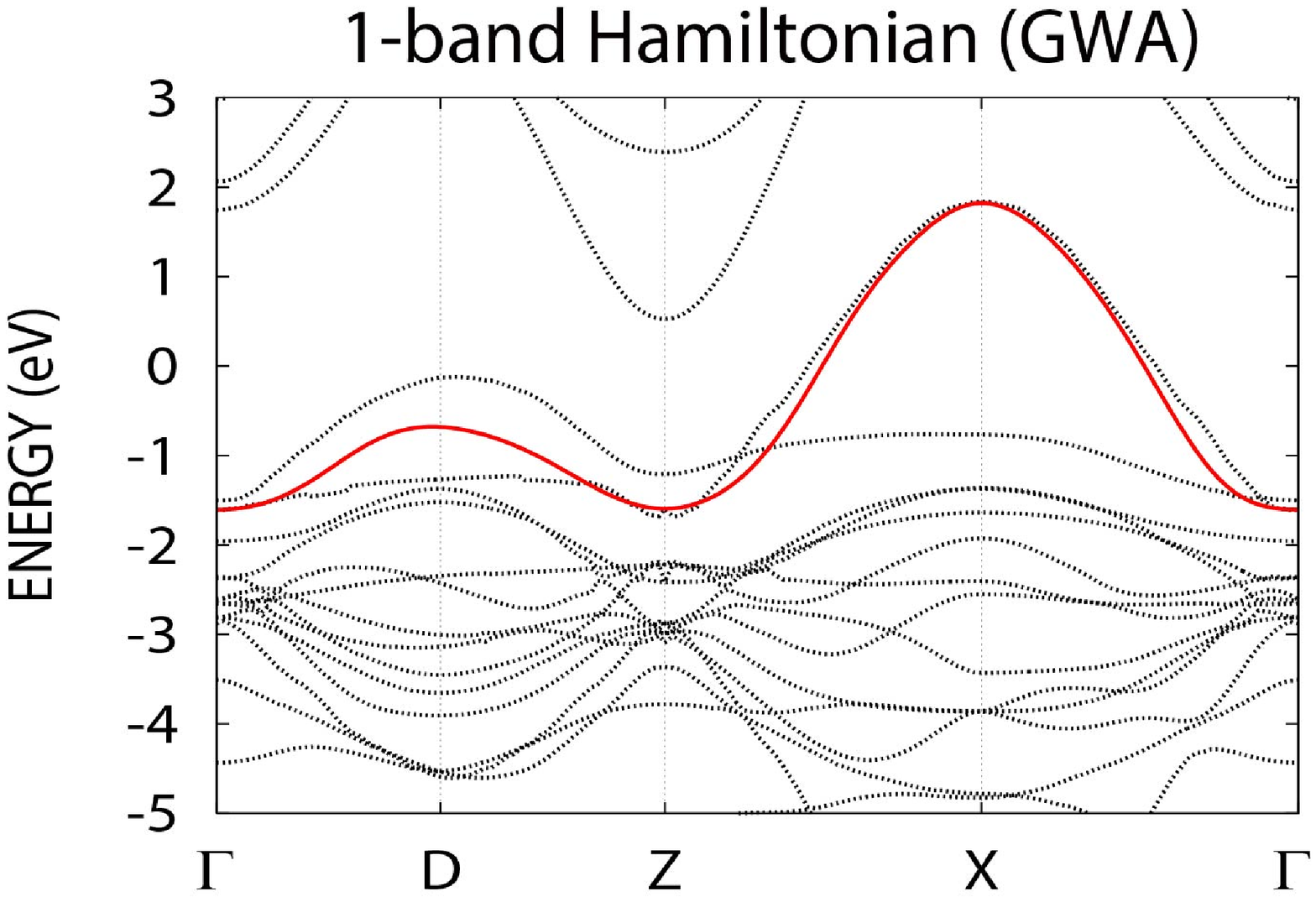} 
\caption{(Color online) Electronic band structure of one-band Hamiltonian in the GWA originating from the Cu $d_{x^2-y^2}$ Wannier orbital for La$_2$CuO$_4$.
The zero energy corresponds to the Fermi level. 
For comparison, the $17$ band structures near the Fermi level in the GWA is also given (black dotted line).
}
\label{bndLaGWwan1}
\end{figure}

\tr{We show the band structure, and parameters for the one-band Hamiltonian in Fig.~\ref{bndLaGWwan1} and Table \ref{paraLa1}, respectively. }

\begin{table*}[h] 
\caption{
Transfer integral and effective interaction of one-band Hamiltonian for La$_2$CuO$_4$ (in eV).
The notations are the same as Table~\ref{paraHg1}. 
}
\ 
\label{paraLa1} 
\begin{tabular}{c|c|c|c|c|c|c|c} 
\hline \hline \\ [-8pt]
$t $(GWA)   &     $(0,0,0)$  &   $(1,0,0)$ &     $(1,1,0)$ &   $(2,0,0)$     \\ [+1pt]
\hline \\ [-8pt] 
  $x^2-y^2$    & 0.187   & -0.451 &  0.088 &  -0.041  \\
\hline \hline \\ [-8pt]
$t $(cGW)   &      $(0,0,0)$  &   $(1,0,0)$ &     $(1,1,0)$ &      $(2,0,0)$    \\ [+1pt]
\hline \\ [-8pt] 
$x^2-y^2$  & -0.003 & -0.482 &  0.073 & -0.102 \\
\hline \hline \\ [-8pt]  
               &      $v$  &     $U(0)$ &     $v_{\text{n}}$ &    $V_{\text{n}}(0)$ &   $v_{\text{nn}}$  &     $V_{\text{nn}}(0)$    \\ [+1pt]
\hline \\ [-8pt]
$x^2-y^2$   &  18.694  & 4.995 &  4.230  &  1.109 &2.824  & 0.765 \\
\hline
\hline 
\end{tabular} 
\end{table*}

\begin{figure}[h]
\centering 
\includegraphics[clip,width=0.4\textwidth ]{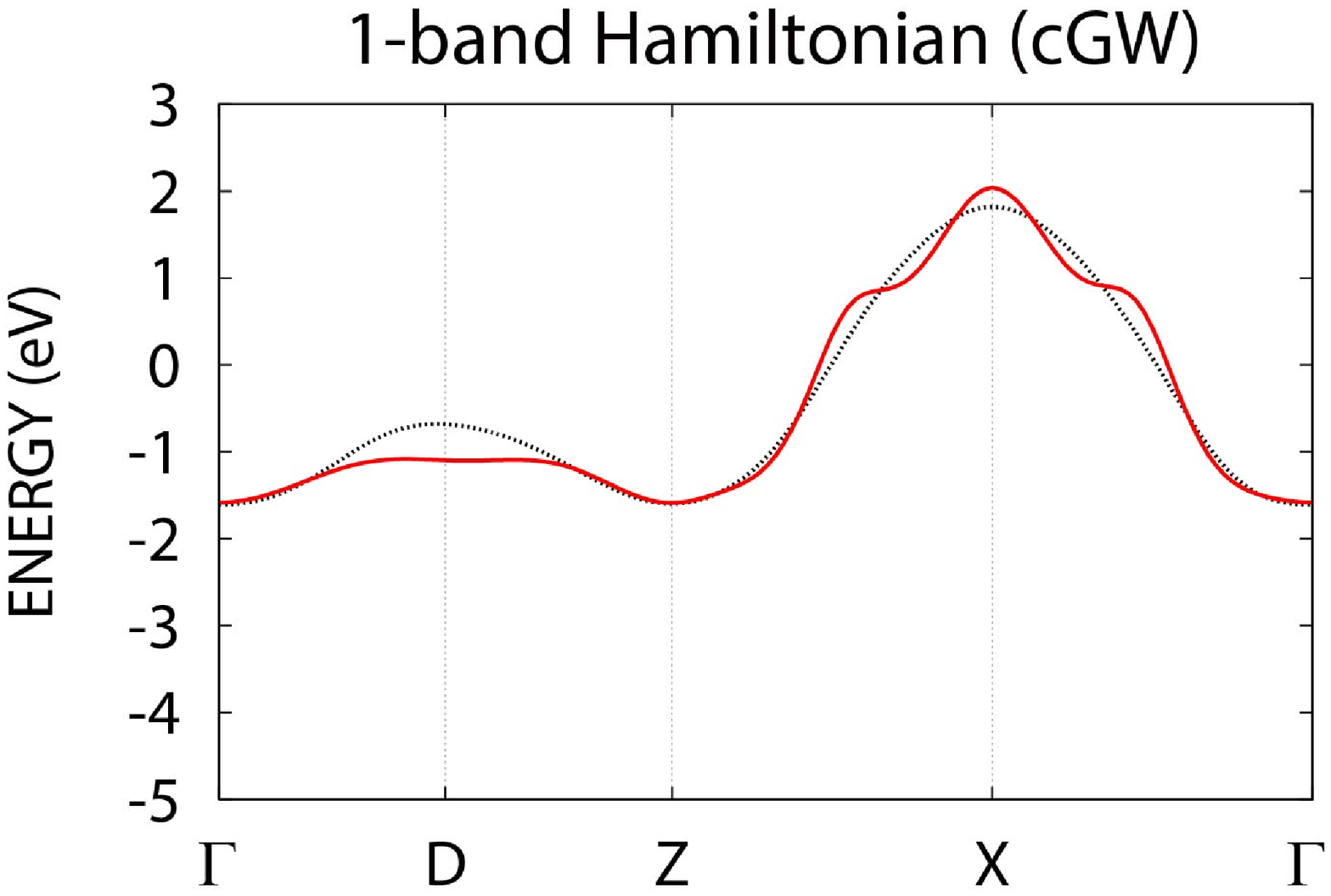} 
\caption{(Color online) Electronic band structure of one-band Hamiltonian in the cGW originating from the Cu $d_{x2-y^2}$ Wannier orbital for La$_2$CuO$_4$.
The zero energy corresponds to the Fermi level. 
For comparison, the band structure in the GWA is also given (black dotted line).
}
\label{bndLaGWcGW1}
\end{figure} 

\subsubsection{three-band Hamiltonian}
\begin{figure}[h]
\centering 
\includegraphics[clip,width=0.4\textwidth ]{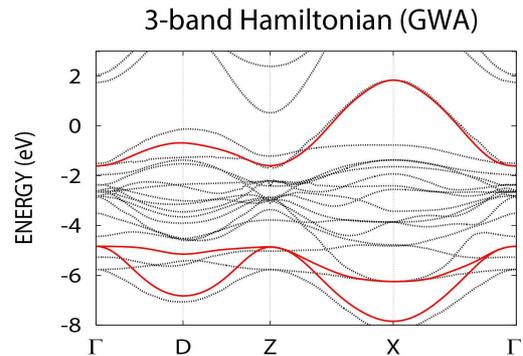} 
\caption{(Color online) Electronic band structure of three-band Hamiltonian in the GWA originating from the Cu $d_{x^2-y^2}$ Wannier orbital for La$_2$CuO$_4$.
The zero energy corresponds to the Fermi level. 
For comparison, the $17$ band structures near the Fermi level in the GWA is also given (black dotted line).
}
\label{bndLaGWwan3}
\end{figure}

\tr{We show the 
\tbb{Wannier function,} 
GWA band structure, cGW+SIC  band structure and parameters for the three-band Hamiltonian in Figs.~
\tbb{\ref{wanLa2}(c),(d),} 
\ref{bndLaGWwan3}, \ref{bndLaGWcGWSIC3} and Table \ref{paraLa3}, respectively. More detailed data can be found in Supplementary Material~\footnotemark[1] including smaller energy parameters.}
\begin{figure}[h]
\centering 
\includegraphics[clip,width=0.4\textwidth ]{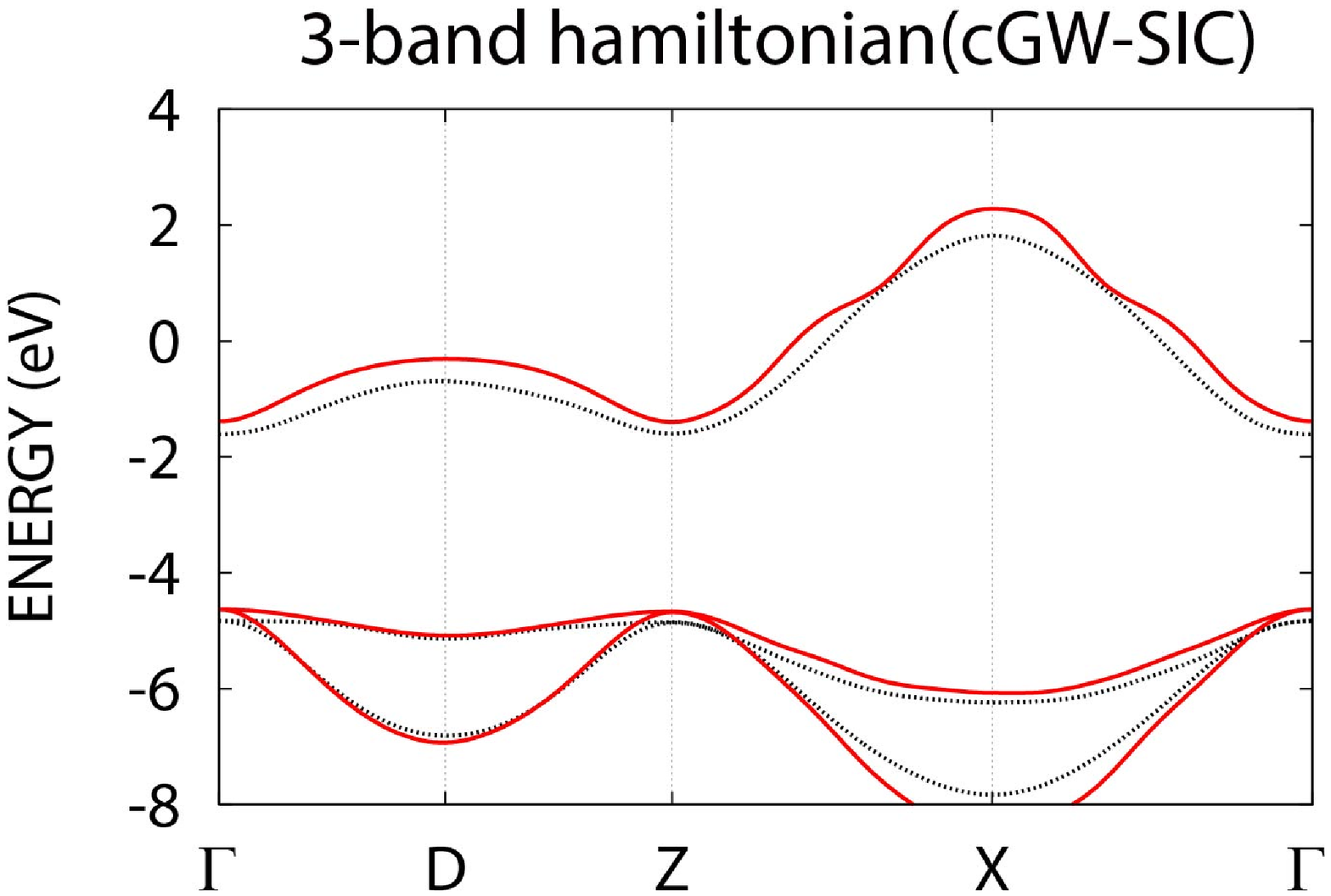} 
\caption{(Color online) Electronic band structures of three-band Hamiltonianin the cGW+SIC originating from the Cu $d_{x2-y^2}$ Wannier orbital for La$_2$CuO$_4$.
The zero energy corresponds to the Fermi level. 
For comparison, the band structure in the GWA is also given (black dotted line).
}
\label{bndLaGWcGWSIC3}
\end{figure} 
\begin{table*}[h] 
\caption{
Transfer integrals and effective interactions for three-band Hamiltonian of La$_2$CuO$_4$ (in eV).
The notations are the same as Table~\ref{paraHg3}.
} 
\label{paraLa3} 
\begin{tabular}{c|ccc|ccc|ccc|ccc|ccc} 
\hline \hline \\ [-8pt]
$t $(GWA)   &       &  $(0,0,0)$  &       &     & $(1,0,0)$ &    &       & $(1,1,0)$ &      &     &  $(2,0,0)$ &     \\ [+1pt]
\hline \\ [-8pt]
      &  $x^2-y^2$ &  $p_1$ &  $p_2$ & $x^2-y^2$ &  $p_1$ &  $p_2$ &  $x^2-y^2$ &  $p_1$ &  $p_2$ & $x^2-y^2$ &  $p_1$ &  $p_2$ \\ 
\hline \\ [-8pt] 
$x^2-y^2$  & -1.743 & -1.399 &  1.399  & -0.010 & -0.012 & -0.042 &  0.013 & -0.006 &  0.006 &  -0.004 & -0.000 & -0.001  \\ 
$p_1$          & -1.399 & -4.657 & -0.659  &   1.399 & 0.120  & 0.659  & -0.042 & 0.041 & -0.000 & 0.012 & -0.002 & -0.000  \\
$p_2$          &  1.399 & -0.659 & -4.657  &  -0.042 & 0.000 & -0.011 &  0.042 &  -0.000 &  0.041   & -0.002 &   0.000 & -0.002    \\
\hline \hline \\ [-8pt]
$t $(cGW-SIC)   &       &  $(0,0,0)$  &       &     & $(1,0,0)$ &    &       & $(1,1,0)$ &      &     &  $(2,0,0)$ &     \\ [+1pt]
\hline \\ [-8pt]
      &  $x^2-y^2$ &  $p_1$ &  $p_2$ & $x^2-y^2$ &  $p_1$ &  $p_2$ &  $x^2-y^2$ &  $p_1$ &  $p_2$ & $x^2-y^2$ &  $p_1$ &  $p_2$ \\ 
\hline \\ [-8pt] 
$x^2-y^2$  & -1.538 & -1.369 &  1.369  & 0.038 & -0.036 & -0.028 &  0.025 & -0.020 &  0.020 & -0.005 &  0.005 &  0.005 \\ 
$p_1$          & -1.369 & -5.237 & -0.753  &  1.369 &  0.189 &  0.754  &  -0.028 &  0.047 &  0.010 & 0.036 & -0.005 &  0.009 \\
$p_2$          &  1.369 & -0.753 & -5.237  &  -0.029 & -0.010 &  0.021 &  0.028 & 0.009 &  0.047  & 0.005 & -0.002 &  0.002   \\
\hline \hline \\ [-8pt]  
   &       &  $v$  &       &     & $U(0)$ &    &       & $J_{v}$ &      &     &  $J(0)$ &     \\ [+1pt]
\hline \\ [-8pt]
      &  $x^2-y^2$ &  $p_1$ &  $p_2$ & $x^2-y^2$ &  $p_1$ &  $p_2$ &  $x^2-y^2$ &  $p_1$ &  $p_2$ & $x^2-y^2$ &  $p_1$ &  $p_2$ \\ 
\hline \\ [-8pt] 
$x^2-y^2$ & 28.784  &  8.246 &   8.246   & 9.612 &  2.680 &   2.680  &            &  0.065 &   0.065  &           &   0.049 &  0.049  \\ 
$p_1$         &   8.246  & 17.777 &  5.501  &  2.680  & 6.128  &  1.861  &  0.065  &            &  0.036  & 0.049  &  -          &  0.019 \\
 $p_2$        &   8.246  &  5.501  & 17.777  & 2.680  &  1.861 &   6.128 &  0.065   & 0.036  &            &  0.049  & 0.019 &         \\
\hline \hline \\ [-8pt]  
       &       &  $v_{\text{n}}$ &    &     & $V_{\text{n}}(0)$ &    &       & $v_{\text{nn}}$  &      &     &  $V_{\text{nn}}(0)$ &     \\ [+1pt]
\hline \\ [-8pt] 
      &  $x^2-y^2$ &  $p_1$ &  $p_2$ & $x^2-y^2$ &  $p_1$ &  $p_2$ &  $x^2-y^2$ &  $p_1$ &  $p_2$ & $x^2-y^2$ &  $p_1$ &  $p_2$ \\ 
\hline \\ [-8pt] 
$x^2-y^2$ &   3.897 &  8.246  &  3.441  & 1.511  &  2.680  & 1.353   &  2.779  &  3.441  &  3.441 &  1.208  & 1.354  & 1.354  \\
$p_1$         &   2.656  & 4.002  &  2.502  &  1.199  &  1.503  &  1.156  &  2.241  & 2.770   & 2.502  &  1.104  & 1.217 &  1.157 \\
$p_2$        &   3.441  & 5.501  &  3.727   &  1.354  &  1.862  &  1.394  &  2.241  &  2.502 &  2.770  &  1.104  & 1.157  & 1.217 \\
\hline \\ [-8pt]
occ.(GWA)      &  $x^2-y^2$ &  $p_1$ &  $p_2$  \\ 
\hline \\ [-8pt] 
                     & 1.350 & 1.825  & 1.825   \\
\hline
\hline 
\end{tabular} 
\end{table*}

%

\section{Discussion}
\subsection{Comparison of the parameters for the La and Hg compounds} \label{DiscussionA}

Main difference of the {\it ab initio} effective Hamiltonians in between the Hg and La compounds arises from the nature of the antibonding band formed from Cu $x^2-y^2$ orbital and two in-plane O $2p_{\sigma}$ orbitals in relation to the band mainly originating from Cu $3z^2-r^2$ orbital hybridizing with the apex oxygen $p_z$ orbital. 

The first difference comes from the level difference $\Delta_{dp}$ between the  
 Cu $x^2-y^2$ orbital and two O $2p_{\sigma}$ orbitals \tg{in the three-band Hamiltonian}.  For the Hg compound, $\Delta_{dp}\sim 2.4$eV while \tg{$\sim 3.7$} eV for the La compound. This difference makes the hybridization between Cu $x^2-y^2$ orbital and two O $2p_{\sigma}$ orbitals substantially larger for the Hg compound. Consequently, the antibonding Wannier orbital constructed from the $x^2-y^2$ and $2p_{\sigma}$ atomic orbitals are more extended to the atomic O position.  This more covalent nature of the Hg compound causes the effective interaction for the Hg compound smaller than the La compound in the one- and two-band Hamiltonians because of the extended Wannier orbital and the stronger screening. This is reflected in the onsite effective interaction of the $x^2-y^2$ antibonding band, which is $U\sim 4.5$ (4.4) eV for the 2-band (1-band) effective Hamiltonian of the Hg compound in comparison to $U\sim 5.5$ (5.0) eV for the La compound. 

\tor{The difference also comes from the fact that the conduction bands of HgBa$_2$CuO$_4$ originating from the $s$-orbitals of the Hg and Ba atoms have wide band widths. 
It is hybridized with 17 bands of the $dp$ orbitals around the Fermi level,
and cross to the bottom of the 17 bands at the $\Gamma $ point (Fig.~\ref{bndHgGW}).
On the other hand, since the La$_2$CuO$_4$ does not have cations that effectively screens the target orbitals,
it shows a stronger interaction than the HgBa$_2$CuO$_4$.}
The poorer screening also makes the effective interaction $U$ for the $3z^2-r^2$ band of the two-band Hamiltonian larger (\tc{$\sim 8.0$} eV) for the La compound than the Hg compound (6.9 eV). 

\trr{Another difference could come from the existence of La $4f$ bands that requires an additional treatment of GW specifically for the $4f$ bands although they do not belong to the 17 bands.  On the physical grounds, we expect that although La $4f$ is located close to the Fermi level in LDA, the correlation effect on the $4f$ bands pushes up the $4f$ levels and the screening effects from the $4f$ bands becomes small, which makes the distinction from the Hg compound less serious in this aspect. This contributes to preserve the larger effective interaction for the La compounds.}

\tg{The level difference of the antibonding $x^2-y^2$ band and the $3z^2-r^2$ band is slightly smaller for the La compound (\tg{$\sim 3.7$} eV) in comparison to the Hg compound (\tg{$\sim 4.0$} eV). Together with the larger $U$,  the La compound has a heavier entanglement of the two bands. Therefore, 
it is plausible that the $3z^2-r^2$ orbital is substantially involved in the low-energy physics near the Fermi level and careful comparisons between the two-band and one-band Hamiltonians would be required for the La compound.}
\tg{The strong entanglement that depends on the momentum in the La compound revealed already in the DFT level makes the one-band treatment of the La compound questionable. \trr{In the DFT level, the two $e_g$ bands strongly hybridize around the D point in the Brillouin zone.} At least it is necessary to confirm the similarity to the solution of the two-band Hamiltonian to justify the one-band Hamiltonian treatment after solving and comparing the both.} 

\tb{The one-body parameters show \tg{another} substantial difference: 
\tg{Although the nearest neighbor transfer of $d_{x^2-y^2}$ orbital, $t_{x^2-y^2}$ for the 1-band (2-band) Hamiltonians is similar (  \tg{-0.46 (-0.43)} eV for the Hg compound and  \tg{ -0.48 (-0.39)} eV for the La compound), the next nearest neighbor transfer $t'_{x^2-y^2}$ shows a substantial difference ( 0.12 \tg{(0.10)} eV for the Hg compound and \tg{0.07 (0.14)} eV for the La compound).   
The ratio $|t'_{x^2-y^2}/t_{x^2-y^2}|$ between the nearest and next-nearest neighbor transfers of the $3d_{x^2-y^2}$ orbital is then around \tg{0.26 (0.24)} for the one-band (two-band) Hamiltonians of the Hg compound, while it is \tg{0.15 (0.35)} for the La compound.  
\tg{A large difference in $t'$ between the two- and one-band parameters of La$_2$CuO$_4$ is ascribed to the fact that the $x^2-y^2$ and $3z^2-r^2$ orbitals in the two-band Hamiltonian entangles and mixes strongly in the one-band Hamiltonian especially in the D point of the Brillouin zone.}
\tg{The present $|t'_{x^2-y^2}/t_{x^2-y^2}|$ for the one-band Hamiltonian shows substantially larger value for the Hg compound than the La compound. This tendency is qualitatively similar to those in Ref.~\onlinecite{Andersen}, where $|t'_{x^2-y^2}/t_{x^2-y^2}|\gtrsim 0.3$ for the Hg compound and $|t'_{x^2-y^2}/t_{x^2-y^2}|\lesssim 0.2$ for the La compound at the LDA level, while the ratios for the two compounds are substantially smaller in the estimation of Ref. \onlinecite{sakakibara2010}. }}}

\tg{Moreover the third neighbor transfer has a non-negligible value $\sim 0.048$ eV for the Hg compound while it is small $\sim0.002$ eV for the La compound. }

\tc{Since the hybridization between the Cu $3d_{x^2-y^2}$ and the oxygen $2p_{\sigma}$ orbitals are strong, we have large splitting of the antibonding band from the nonbonding and bonding orbitals. This is the basis of justifying the one- or two-band Hamiltonians rather than the three-band form~\cite{ZhangRice}.  However, since the interaction scale is not absolutely smaller than the splitting, it is conceivable that the effect of the charge fluctuation between the Cu $3d_{x^2-y^2}$ and the oxygen $2p_{\sigma}$ orbitals appears in some physical quantities as first pointed out in~Ref.~\onlinecite{Varma}.  The present three-band Hamiltonians will serve for the purpose of examining the relevance of dynamical $3d_{x^2-y}$-$2p_{\sigma}$ fluctuations from the comparisons with the one-band results based on first-principles and realistic analyses. This is especially important for the Hg compound because $\Delta_{dp}$ is smaller.}

\tb{We believe that the substantial differences revealed above must lead to various differences in physical properties, particularly in the difference in the critical temperature. This paper provides a starting point for understanding such differences. By solving the effective Hamiltonians in future studies, \tr{consequences} of the differences will be elucidated.  Especially, it was shown~\cite{misawa2014} that the phase separation is enhanced 
\tg{if $|t'/t|$ becomes small for the Hubbard model. The phase separation is also enhanced }
for larger $U/t$ in the Hubbard model. Then in the present realistic Hamiltonians, \tg{these two differences may cooperatively} enhance the charge inhomogeneity of the La compound in comparison to the Hg compound.  This is consistent with the experimental observation that the La compound has a stronger tendency to the stripe and charge inhomogeneities.  Stronger effective attraction of carriers is required to reach high $T_{\rm c}$, while this is a double edged sword, because it also drives the inhomogeneity including stripes and charge orders~\cite{MisawaSciAdv}.  The relation of the inhomogeneity and the critical temperature and ways to enhance $T_{\rm c}$ by suppressing the inhomogeneity is an interesting future issue .  } 

\tred{The one-band Hamiltonian is justified when the Hilbert subspace for the antibonding band is essentially retained even after taking effective Coulomb interactions into account at and around the Mott insulator. The reconstruction that invalidates the one-band description will be negligible  when the level splitting $\mu_{\rm ab}-\mu_{\rm b}$ between the antibonding orbital and the bonding (or nonbonding) orbitals is larger than the difference $U_{\rm b}-U'_{\rm bab}$ between the onsite effective Coulomb repulsion within the bonding or nonbonding oribtal ($U_{\rm b}$ or $U_{\rm nb}$) and the onsite repulsion $U'_{\rm bab}$ between an antibonding electron and a bonding (or nonbonding) electron. The level splittings $\mu_{\rm ab}-\mu_{\rm b}$ is 4 eV or larger as one sees in Figs.\ref{bndHgGWcGWSIC3} and \ref{bndLaGWcGWSIC3}, while $U_{\rm b}-U'_{\rm bab}$ may not exceed 4eV.  Namely, the energy level of the upper Hubbard band for the bonding or nonbonding orbital may be lower than the energy level of the lower Hubbard for the antibonding band.  Hence the doped hole is expected to preserve the character of the antibonding orbital.This is one reasoning for the justification of the one-band Hamiltonian and the description by Zhang-Rice singlet~\cite{ZhangRice}. Since the energy differences discussed above is not overwhelmingly large, uncertainties remain. Therefore, the final answer to the validity of the description by one-band hamoltonians will be obtained after solving the Hamiltonian in the future.}

\section{Summary}
\tr{ We have derived {\it ab initio} low-energy effective Hamiltonians for La$_2$CuO$_4$ and HgBa$_2$CuO$_4$, on the basis of the multi-scale {\it ab initio} scheme for correlated electrons (MACE).  Among MACE, we have employed a refined scheme to eliminate the double counting of electron correlations arising from the DFT and the procedure of solving the presently derived Hamiltonians by low-energy solvers afterwards. Three different effective Hamiltonians are derived: 1)  one-band Hamiltonian for the antibonding orbital generated from strongly hybridized Cu $3d$ $x^2-y^2$ and O $2p_{\sigma}$ orbitals  
2) two-band Hamiltonian constructed from the Cu $3d$ $3z^2-r^2$ orbital in addition to the above antibonding  $3d$ $x^2-y^2$ orbital.  For the two-band Hamiltonians, we have prepared two options. In the first choice, the Cu $3d_{3z^2-r^2}$ orbital is treated as the atomic-like and the direct contribution from the oxygen $2p_z$ orbital is treated as the eliminated high-energy part. In the second choice, the $2p_z$ orbital hybridizing with the Cu $3d_{3z^2-r^2}$ orbital is taken into account in the low energy Hamiltonian. Then the antibonding orbital constructed from the Cu $3d_{3z^2-r^2}$ and  the $2p_z$ orbitals constitutes one of the two bands in the effective Hamiltonian. The two choices give substantially different effective interactions for the band involving the Cu $3d_{3z^2-r^2}$ orbitals.  After solving the effective Hamiltonian,  however, we expect that the two choices give similar results, if the  Cu $3d_{3z^2-r^2}$ orbitals play minor roles in low-energy thermodynamic properties at the scale of the room temperature. If the $3d_{3z^2-r^2}$ orbitals play roles, careful comparisons between two choices are required.
3) Three-band Hamiltonian consisting mainly of Cu $3d$ $x^2-y^2$ orbitals and two O $2p_{\sigma}$ orbitals. }  

\tr{Main differences between the Hamiltonians for  La$_2$CuO$_4$ and HgBa$_2$CuO$_4$ are summarized in the following three points.   i) The two oxygen $2p_{\sigma}$} orbitals are farther (\tg{$\sim 3.7$} eV) below from the Cu $d_{x^2-y^2}$ orbital for the La compound than the Hg compound ($\sim 2.4$ eV),  which makes effective onsite Coulomb interaction $U$ for the antibonding  $d_{x^2-y^2}$-$2p_{\sigma}$ band larger for the La compound (5.5 (5.0) eV) than the Hg compound (4.5 (4.0) eV) in the two-band (one-band) Hamiltonians. The difference is also enhanced by the screening by the $s$ band originating from the cations (Hg and Ba), which is located closer to the CuO$_2$ plane and has energy closer to the Fermi level than the La cation $s$ band.  ii) The ratio of the second-neighbor to the nearest transfer $t'/t$ is also substantially different (\tg{0.26 for the Hg and 0.15 for the La compound for the one-band Hamiltonian}). 
\tg{iii) The level difference of the bands mainly consisting of the copper $d_{x^2-y^2}$ from the $d_{3z^2-r^2}$ orbitals is slightly larger for the Hg compound (\tg{$\sim 4.0$} eV) than the La compound (\tg{$\sim 3.7 $} eV). Combined with the larger onsite interaction, the La compound has heavier entanglement of the two bands for the La compound.  Therefore, the 1-band Hamiltonian could be insufficient in representing some aspects of the La compound. }

\tr{The effective Hamiltonians obtained in the present study serve as platforms of future studies aiming at accurately solving the low-energy effective Hamiltonians beyond the density functional theory.  Further studies on physics of superconductivity on the cuprates based on the present {\it ab initio} effective Hamiltonians are highly desirable.  The present study may also promote future design of higher $T_{\rm c}$ based on the first principles approach, which is another intriguing future subject. 
}
\begin{acknowledgments}
The authors thank Kosuke Miyatani for his help and contribution in the initial stage.  They are also indebted to Takashi Miyake for his advice.  The authors also acknowledge Terumasa Tadano, Takahiro Ohgoe, Yusuke Nomura and Kota Ido for useful discussions.
This work was financially supported by a Grant-in-Aid for Scientific Research (No. 22104010, No. 16H06345 and No.16K17746) from Ministry of Education, Culture, Sports, Science and Technology, Japan.  
This work was also supported in part by MEXT as a social and scientific priority issue (Creation of new functional devices and high-performance 
materials to support next-generation industries; CDMSI) to be tackled by using post-K computer.
The authors thank the Supercomputer Center, the Institute for Solid State Physics, the University of Tokyo for the facilities. 
We thank the computational resources of the K computer provided by the RIKEN Advanced Institute for Computational Science through the HPCI System Research project (hp150173, hp150211, hp160201,hp170263) supported by Ministry of Education, Culture, Sports, Science, and Technology, Japan. 
TM is supported by Building of Consortia for the Development of 
Human Resources in Science and Technology from the MEXT of Japan.

\end{acknowledgments}
%
%
\appendix
\section{Two-band Hamiltonian for \tor{La$_2$CuO$_4$} with antibonding $3d_{z^2-r^2}-2p_z$ orbital}
\label{AppendixA} 
\tg{Here we present two-band Hamiltonian parameters in Table~\ref{paraLa2_2}, which is an alternative to Table \ref{paraLa2}. 
One of the two bands is constructed from the antibonding band consisting of  the copper $3d_{z^2-r^2}$ orbital and the apex oxygen $2p_z$ orbital. The other band is the antibonding band consisting of  the copper $3d_{x^2-y^2}$ orbital and the inplane oxygen $2p_{\sigma}$ orbitals. }

\begin{table*}[h] 
\caption{
Transfer integral and effective interaction in two-band Hamiltonian for La$_2$CuO$_4$ (in eV) \tg{where one of the two bands is constructed from the antibonding band of the copper $3z^2-r^2$ and apex oxygen $p_z$ orbitals.}
Notations are the same as Table~\ref{paraHg2}.
}
\ 
\label{paraLa2_2} 
\begin{tabular}{c|cc|cc|cc|cc|cc} 
\hline \hline \\ [-8pt]
$t $(GWA)   &       &  $(0,0,0)$  &      & $(1,0,0)$ &       & $(1,1,0)$ &      &  $(2,0,0)$     \\ [+1pt]
\hline \\ [-8pt]
      &  $3z^2-r^2 $ &  $x^2-y^2 $  &  $3z^2-r^2 $ &  $x^2-y^2 $ &  $3z^2-r^2 $ &  $x^2-y^2 $ &  $3z^2-r^2 $ &  $x^2-y^2 $ \\ 
\hline \\ [-8pt] 
$3z^2-r^2 $  & -0.958  & 0.000 &  -0.047  & 0.151 &   -0.035 &  0.000 & 0.019 & 0.007 \\
$x^2-y^2 $  &  0.000  &  -0.012  & 0.151 & -0.448 & 0.000  & 0.089 &  0.007 & -0.043 \\
\hline \hline \\ [-8pt]
$t $(cGW-SIC)   &       &  $(0,0,0)$  &      & $(1,0,0)$ &        & $(1,1,0)$ &        &  $(2,0,0)$     \\ [+1pt]
\hline \\ [-8pt]
      &  $3z^2-r^2 $ &  $x^2-y^2 $  &  $3z^2-r^2 $ &  $x^2-y^2 $ &  $3z^2-r^2 $ &  $x^2-y^2 $ &  $3z^2-r^2 $ &  $x^2-y^2 $ \\ 
\hline \\ [-8pt] 
$3z^2-r^2 $  &  -0.212 & 0.000 &  -0.038 & 0.086 &  0.009 &  0.000 & -0.017 & 0.012 \\
$x^2-y^2 $   &  0.000  & 0.138 &  0.086 & -0.389 & 0.000 &  0.143&  0.012 & 0.001 \\
\hline \hline \\ [-8pt]  
   &       &  $v$  &      & $U(0)$ &     & $J_{v}$ &       &  $J(0)$      \\ [+1pt]
\hline \\ [-8pt]
       &  $3z^2-r^2 $ &  $x^2-y^2 $  &  $3z^2-r^2 $ &  $x^2-y^2 $ &  $3z^2-r^2 $ &  $x^2-y^2 $ &  $3z^2-r^2 $ &  $x^2-y^2 $ \\ 
\hline \\ [-8pt] 
$3z^2-r^2 $   & 16.172 & 15.558 &  4.878  & 3.826 &             &   0.673   &               &  0.550 \\
$x^2-y^2 $  & 15.558 & 18.505 & 3.826  & 5.320 & 0.673  &                &  0.550   &             \\ 
\hline \hline \\ [-8pt]  
       &       &  $v_{\text{n}}$ &    & $V_{\text{n}}(0)$ &       & $v_{\text{nn}}$  &       &  $V_{\text{nn}}(0)$     \\ [+1pt]
\hline \\ [-8pt] 
      &  $3z^2-r^2 $ &  $x^2-y^2 $  &  $3z^2-r^2 $ &  $x^2-y^2 $ &  $3z^2-r^2 $ &  $x^2-y^2 $ &  $3z^2-r^2 $ &  $x^2-y^2 $ \\ 
\hline \\ [-8pt] 
$3z^2-r^2 $  & 3.452  & 3.775  & 1.325  & 1.411 & 2.584  & 2.684 & 1.145  &   1.164 \\
$x^2-y^2 $  &   3.775 &  4.240 & 1.411  & 1.539 &  2.684 &  2.823 & 1.164   &  1.193 \\
\hline \\ [-8pt]
occ.(GWA)      &  $3z^2-r^2 $ &  $x^2-y^2 $  \\ 
\hline \\ [-8pt] 
                     & 1.949  &  1.051   \\
\hline
\hline 
\end{tabular} 
\end{table*} 
\bibliographystyle{apsrev}
\bibliography{Hirayama_cuprate_Ref.}
\end{document}


\title{Supplementary material for {\it ab initio} effective Hamiltonians for cuprate superconductors}
  \author{Motoaki Hirayama$^{1)}$, Youhei Yamaji$^{2)}$, Takahiro Misawa$^{3)}$ and Masatoshi Imada$^{2)}$}
  \affiliation{$^{1)}$Department of Physics, Tokyo Institute of Technology, Japan}
  \affiliation{$^{2)}$Department of Applied Physics, University of Tokyo, 7-3-1 Hongo, Bunkyo-ku, Tokyo 113-8656, Japan}
    \affiliation{$^{3)}$Institute for Solid State Physics, University of Tokyo, Kashiwanoha, Kashiwa, Chiba, Japan}
\maketitle    
%
%
\begin{widetext}

\section{Spread of Wannier function}

We show the spread of the maximally localized Wannier functions (MLWFs) for HgBa$_2$CuO$_4$ and  La$_2$CuO$_4$  in Table~\ref{spread}.

\begin{table}[h!] 
\caption{Spread of the Wannier functions (in \AA $^{2}$) defined by quadratic extent.
The upper/lower row shows the values of the two-/three-band model.
} 
\
\begin{tabular}{c|cc}
\hline \hline \\ [-8pt]  
 Orbital            & HgBa$_2$CuO$_4$ & La$_2$CuO$_4$    \\ [+1pt]
\hline \\ [-8pt] 
Cu  $3z^2-r^2$                & 1.13 & 0.68   \\ 
Cu $x^2-y^2$ anti-bonding       & 2.68 &  2.14   \\
\hline    \\
Cu $x^2-y^2$                            & 0.56 & 0.52   \\
O $2p$                                       & 1.78 & 1.24  \\
\hline \hline 
\end{tabular}
\label{spread} 
\end{table}

%
%

%
%

\section{Details of hamiltonans}

In this supplementary material, we list up the whole parameters including relatively small one-body parameters up to the relative unit-cell distance (3,3,0). Beyond (3,3,0)  all the one-body parameters are below 10 meV.  We also list up two-body parameters up to the distance (3,3,0). Interactions for further neighbor unit-cell pairs very well follows $1/r$ dependence inferred from the list.
One-body parameters for the two-band hamiltonian of HgBa$_2$CuO$_4$ are listed in Table \ref{paraHg2all} and the interaction parameters are given in Tables~
S.2, S.3, and S.4.
The one-band hamiltonian parameters are listed in Tables S.5 and S.6.
in the same way.  The three-band hamiltonian parameters are in  Tables 
S.7, S.8, S.9, S.10, S.11, and S.12.
The hamiltonian parameters for La$_2$CuO$_4$ are given in the same order in Tables S.13-S.23. Note that the unit cell of  La$_2$CuO$_4$ has two copper atoms in the $z$ direction.

\begin{table*}[h] 
\caption{
Transfer integrals and onsite potentials in the cGW-SIC for the two-band hamiltonian of HgBa$_2$CuO$_4$ (in eV).
The inter-layer hopping except for that in $(0,0,1)$ is omitted because its energy scale is under 10 meV.
}
\ 
\label{paraHg2all} 
 
\end{center} 
\end{table} 
\begin{table*}[h] 
\label{paraHg3all} 
\begin{center}
\caption{
Transfer integrals in the cGW-SIC for three-band hamiltonian of HgBa$_2$CuO$_4$ (in eV).
The inter-layer hopping is omitted because its energy scale is under 10 meV.
}
\ 
%
 
\end{tabular} 
\end{center}
\end{minipage}
\begin{minipage}{1.0\hsize} 
\label{WrHg2all_interlayer} 
\begin{center}
\caption{
Interlayer effective interactions in the cGW-SIC for three-band hamiltonian of HgBa$_2$CuO$_4$ (in eV).
Notations are the same as Table S8.
}
\ 
 
\end{center}
\end{table*}



\begin{table*}[h] 
\caption{
Transfer integrals in the cGW-SIC for two-band hamiltonian of La$_2$CuO$_4$ (in eV).
The inter-layer hopping is omitted because its energy scale is under 10 meV.
}
\ 
\label{paraLa2all} 
\begin{tabular}{c|cc|cc|cc|cc|cc} 
\hline \hline \\ [-8pt]
$t $(cGW-SIC)   &       &  $(0,0,0)$  &      & $(1,0,0)$ &        & $(1,1,0)$ &        &  $(2,0,0)$     \\ [+1pt]
\hline \\ [-8pt]
      &  $3z^2-r^2 $ &  $x^2-y^2 $  &  $3z^2-r^2 $ &  $x^2-y^2 $ &  $3z^2-r^2 $ &  $x^2-y^2 $ &  $3z^2-r^2 $ &  $x^2-y^2 $ \\ 
\hline \\ [-8pt] 
$3z^2-r^2 $  &  -3.426 & 0.000 &  -0.008 & 0.057 &  -0.013 &  0.000 & 0.006 & 0.009 \\
$x^2-y^2 $   &  0.000  & 0.313 &  0.057 & -0.389 & 0.000 &  0.136 &  0.009 & 0.003 \\
\hline \hline \\ [-8pt]
   &       &  $(2,1,0)$  &      & $(2,2,0)$ &        & $(3,0,0)$ &        &  $(3,1,0)$     \\ [+1pt]
\hline \\ [-8pt]
      &  $3z^2-r^2 $ &  $x^2-y^2 $  &  $3z^2-r^2 $ &  $x^2-y^2 $ &  $3z^2-r^2 $ &  $x^2-y^2 $ &  $3z^2-r^2 $ &  $x^2-y^2 $ \\ 
\hline \\ [-8pt] 
$3z^2-r^2 $  &  -0.004 &  0.001 & 0.002 &  0.000  & 0.003  & 0.001 & -0.002 & 0.001 \\
$x^2-y^2 $   & 0.001 & -0.020 & 0.000 &  0.000 & 0.001 &  0.012 &  0.001 &   -0.009 \\
\hline \hline \\ [-8pt]
   &       &  $(3,2,0)$  &      & $(3,3,0)$     \\ [+1pt]
\hline \\ [-8pt]
      &  $3z^2-r^2 $ &  $x^2-y^2 $  &  $3z^2-r^2 $ &  $x^2-y^2 $  \\ 
\hline \\ [-8pt] 
$3z^2-r^2 $  &  0.000  0.001 & 0.000  & 0.0000    \\
$x^2-y^2 $   & 0.001  0.002 & 0.000 &  0.002  \\
\hline
\hline 
\end{tabular} 
\end{table*}

\begin{table*}[h] 
\caption{
Effective interactions in the cGW-SIC for two-band hamiltonian of La$_2$CuO$_4$ (in eV).
Notations are the same as Table S2.
}
\ 
\label{WrLa2all} 
\begin{tabular}{c|c} 
\hline \hline \\ [-8pt]
   $\bm{R}_{i_1}$ \ \ $\ell_1$ \ \ $\bm{R}_{i_2}$ \ \ $\ell_2$ \ \ $\bm{R}_{i_3}$ \ \ $\ell_3$ \ \ $\bm{R}_{i_4}$ \ \ $\ell_4$ & $v$ \ \ \ \  \ \ $W^r$     \\ [+1pt]
  $(0,0,0)$  1  $(0,0,0)$  1  $(0,0,0)$  1  $(0,0,0)$  1 & 26.091   7.993   \\
  $(0,0,0)$  1  $(0,0,0)$  1  $(0,0,0)$  2  $(0,0,0)$  2 & 20.037   4.906   \\
  $(0,0,0)$  1  $(0,0,0)$  2  $(0,0,0)$  1  $(0,0,0)$  2 & 0.874   0.793   \\
  $(0,0,0)$  1  $(0,0,0)$  2  $(0,0,0)$  2  $(0,0,0)$  1 &  0.874   0.793   \\
  $(0,0,0)$  2  $(0,0,0)$  1  $(0,0,0)$  1  $(0,0,0)$  2 & 0.874   0.793   \\
  $(0,0,0)$  2  $(0,0,0)$  1  $(0,0,0)$  2  $(0,0,0)$  1 &  0.874   0.793   \\
\hline     
\hline 
\end{tabular} 
\end{table*} 

\begin{table*}[h] 
\label{WrLa2all_interlayer} 
\caption{
Effective interactions in the cGW-SIC for two-band hamiltonian of La$_2$CuO$_4$ (in eV).
Notations are the same as Table S2.
}
\ 
\begin{tabular}{c|c} 
\hline \hline \\ [-8pt]
  $(0,0,0)$  1  $(0,0,0)$  1  $(3,3,0)$  1  $(3,3,0)$  1  & 1.449   0.931   \\
  $(0,0,0)$  1  $(0,0,0)$  1  $(3,3,0)$  2  $(3,3,0)$  2  & 1.456   0.932   \\
  $(0,0,0)$  2  $(0,0,0)$  2  $(3,3,0)$  1  $(3,3,0)$  1  & 1.456   0.932   \\
  $(0,0,0)$  2  $(0,0,0)$  2  $(3,3,0)$  2  $(3,3,0)$  2  & 1.463   0.933   \\
  $(0,0,0)$  1  $(0,0,0)$  1  $(0.5,0.5,1)$  1  $(0.5,0.5,1)$  1 &  2.101   1.091   \\
  $(0,0,0)$  1  $(0,0,0)$  1  $(0.5,0.5,1)$  2  $(0.5,0.5,1)$  2 &  2.066   1.080   \\
  $(0,0,0)$  2  $(0,0,0)$  2  $(0.5,0.5,1)$  1  $(0.5,0.5,1)$  1 &  2.066   1.080   \\
\hline    
\hline 
\end{tabular} 
\end{table*} 

\begin{table*}[h] 
\label{WrLa2all_offdiagonal} 
\caption{
Effective interactions in the cGW-SIC for two-band hamiltonian of La$_2$CuO$_4$ (in eV).
Notations are the same as Table S2.
}
\ 
\begin{tabular}{c|c} 
\hline \hline \\ [-8pt]
 $(-1,0,0)$  2  $(0,0,0)$  2  $(0,0,0)$  1  $(0,0,0)$  1  & 0.357   0.074    \\
 $(-1,0,0)$  2  $(0,0,0)$  2  $(0,0,0)$  2  $(0,0,0)$  2  & 0.277   0.077   \\
 $(-1,0,0)$  2  $(0,0,0)$  2 $(-1,0,0)$  1 $(-1,0,0)$  1 &  0.357   0.074   \\
 $(-1,0,0)$  2  $(0,0,0)$  2 $(-1,0,0)$  2 $(-1,0,0)$  2 &  0.277   0.077   \\
  $(0,0,0)$  1  $(0,0,0)$  1  $(0,0,0)$  2  $(0,1,0)$  2 &  0.357   0.074   \\
  $(0,0,0)$  2  $(0,0,0)$  2  $(0,0,0)$  2  $(0,1,0)$  2 &  0.277   0.077   \\
  $(0,0,0)$  1  $(0,0,0)$  1  $(0,-1,0)$  2  $(0,0,0)$  2 &  0.357   0.074   \\
  $(0,0,0)$  2  $(0,0,0)$  2  $(0,-1,0)$  2  $(0,0,0)$  2 &  0.277   0.077   \\
  $(0,0,0)$  1  $(0,0,0)$  1  $(0,0,0)$  2  $(0,-1,0)$  2 &  0.357   0.074   \\
  $(0,0,0)$  2  $(0,0,0)$  2  $(0,0,0)$  2  $(0,-1,0)$  2 &  0.277   0.077   \\
  $(0,0,0)$  1  $(0,0,0)$  1  $(0,1,0)$  2  $(0,0,0)$  2 &  0.357   0.074   \\
  $(0,0,0)$  2  $(0,0,0)$  2  $(0,1,0)$  2  $(0,0,0)$  2 &  0.277   0.077   \\
  $(0,0,0)$  1  $(0,0,0)$  1  $(0,0,0)$  2  $(1,0,0)$  2 &  0.357   0.074   \\
  $(0,0,0)$  2  $(0,0,0)$  2  $(0,0,0)$  2  $(1,0,0)$  2 &  0.277   0.077   \\
  $(0,0,0)$  1  $(0,0,0)$  1 $(-1,0,0)$  2  $(0,0,0)$  2 &  0.357   0.074   \\
  $(0,0,0)$  2  $(0,0,0)$  2 $(-1,0,0)$  2  $(0,0,0)$  2 &  0.277   0.077   \\
  $(0,0,0)$  1  $(0,0,0)$  1  $(0,0,0)$  2 $(-1,0,0)$  2 &  0.357   0.074   \\
  $(0,0,0)$  2  $(0,0,0)$  2  $(0,0,0)$  2 $(-1,0,0)$  2 &  0.277   0.077   \\
  $(0,0,0)$  1  $(0,0,0)$  1  $(1,0,0)$  2  $(0,0,0)$  2 &  0.357   0.074   \\
  $(0,0,0)$  2  $(0,0,0)$  2  $(1,0,0)$  2  $(0,0,0)$  2 &  0.277   0.077   \\
\hline      
\hline 
\end{tabular} 
\end{table*}

\begin{table*}[ptb] 
\caption{
Transfer integrals in the cGW for one-band hamiltonian of La$_2$CuO$_4$ (in eV).
The inter-layer hopping is omitted because its energy scale is under 10 meV.
}
\ 
\label{paraLa1all} 
\begin{tabular}{c|c|c|c|c|c|c|c} 
\hline \hline \\ [-8pt]
$t $(cGW)   &      $(0,0,0)$  &   $(1,0,0)$ &     $(1,1,0)$ &      $(2,0,0)$    \\ [+1pt]
\hline \\ [-8pt] 
$x^2-y^2$  & -0.003 & -0.482 &  0.073 & -0.102 \\
\hline \hline \\ [-8pt]
   &      $(2,1,0)$  &   $(2,2,0)$ &     $(3,0,0)$ &      $(3,1,0)$    \\ [+1pt]
\hline \\ [-8pt] 
$x^2-y^2$  & 0.049  & -0.005 &  -0.045 &  0.057 \\
\hline \hline \\ [-8pt]
$t $(cGW)   &      $(3,2,0)$  &   $(3,3,0)$   \\ [+1pt]
\hline \\ [-8pt] 
$x^2-y^2$  & -0.013 & -0.020  \\
\hline
\hline 
\end{tabular} 
\end{table*} 

\begin{table*}[h] 
\caption{
Effective interactions in the cGW-SIC for one-band hamiltonian of La$_2$CuO$_4$ (in eV).
Notations are the same as Table S2.
}
\ 
\label{WrLa1all} 
\begin{tabular}{c|c} 
\hline \hline \\ [-8pt]
   $\bm{R}_{i_1}$ \ \ $\ell_1$ \ \ $\bm{R}_{i_2}$ \ \ $\ell_2$ \ \ $\bm{R}_{i_3}$ \ \ $\ell_3$ \ \ $\bm{R}_{i_4}$ \ \ $\ell_4$ & $v$ \ \ \ \  \ \ $W^r$     \\ [+1pt]
    $(0,0,0)$  1  $(0,0,0)$  1  $(0,0,0)$  1  $(0,0,0)$  1 & 18.694   4.995   \\
  $(0,0,0)$  1  $(0,0,0)$  1  $(1,0,0)$  1  $(1,0,0)$  1 &  4.230   1.109   \\
  $(0,0,0)$  1  $(0,0,0)$  1  $(1,1,0)$  1  $(1,1,0)$  1 &  2.824   0.765   \\
  $(0,0,0)$  1  $(0,0,0)$  1  $(2,0,0)$  1  $(2,0,0)$  1 &  2.135   0.623   \\
  $(0,0,0)$  1  $(0,0,0)$  1  $(2,1,0)$  1  $(2,1,0)$  1 &  1.930   0.587   \\
  $(0,0,0)$  1  $(0,0,0)$  1  $(2,2,0)$  1  $(2,2,0)$  1 &  1.644   0.534   \\
\hline      
\hline 
\end{tabular} 
\end{table*}

\begin{table*}[h] 
\caption{
Transfer integrals in the cGW-SIC for three-band hamiltonian of La$_2$CuO$_4$ (in eV).
The inter-layer hopping is omitted because its energy scale is under 10 meV.
}
\ 
\label{paraLa3all} 
%
 
\end{center}
\end{minipage}

\clearpage

\begin{minipage}{1.0\hsize}
\vspace{-1.2truecm}
\label{WrLa3all_offdiagonal} 
\caption{
Effective interactions in the cGW-SIC for three-band hamiltonian of La$_2$CuO$_4$ (in eV).
Notations are the same as Table S8.
}
\ 
\begin{tabular}{c|c} 
\hline \hline \\ [-8pt]
  $(0,0,0)$  2  $(0,0,0)$  2  $(2,0,0)$  3  $(2,0,0)$  3 &  1.787   1.012   \\ 
  $(0,0,0)$  3  $(0,0,0)$  3  $(2,0,0)$  1  $(2,0,0)$  1 &  2.022   1.060   \\ 
  $(0,0,0)$  3  $(0,0,0)$  3  $(2,0,0)$  2  $(2,0,0)$  2 &  2.502   1.157   \\ 
  $(0,0,0)$  3  $(0,0,0)$  3  $(2,0,0)$  3  $(2,0,0)$  3 &  2.060   1.068   \\ 
  $(0,0,0)$  1  $(0,0,0)$  1  $(2,1,0)$  1  $(2,1,0)$  1 &  1.908   1.038   \\ 
  $(0,0,0)$  1  $(0,0,0)$  1  $(2,1,0)$  2  $(2,1,0)$  2 &  2.241   1.104   \\ 
  $(0,0,0)$  1  $(0,0,0)$  1  $(2,1,0)$  3  $(2,1,0)$  3 &  2.022   1.060   \\ 
  $(0,0,0)$  2  $(0,0,0)$  2  $(2,1,0)$  1  $(2,1,0)$  1 &  1.724   1.001   \\ 
  $(0,0,0)$  2  $(0,0,0)$  2  $(2,1,0)$  2  $(2,1,0)$  2 &  1.906   1.039   \\ 
  $(0,0,0)$  2  $(0,0,0)$  2  $(2,1,0)$  3  $(2,1,0)$  3 &  1.787   1.013   \\ 
  $(0,0,0)$  3  $(0,0,0)$  3  $(2,1,0)$  1  $(2,1,0)$  1 &  1.750   1.008   \\ 
  $(0,0,0)$  3  $(0,0,0)$  3  $(2,1,0)$  2  $(2,1,0)$  2 &  1.953   1.048   \\ 
  $(0,0,0)$  3  $(0,0,0)$  3  $(2,1,0)$  3  $(2,1,0)$  3 &  1.890   1.037   \\ 
  $(0,0,0)$  1  $(0,0,0)$  1  $(2,2,0)$  1  $(2,2,0)$  1 &  1.631   0.985   \\ 
  $(0,0,0)$  1  $(0,0,0)$  1  $(2,2,0)$  2  $(2,2,0)$  2 &  1.750   1.008   \\ 
  $(0,0,0)$  1  $(0,0,0)$  1  $(2,2,0)$  3  $(2,2,0)$  3 &  1.750   1.008   \\ 
  $(0,0,0)$  2  $(0,0,0)$  2  $(2,2,0)$  1  $(2,2,0)$  1 &  1.544   0.967   \\ 
  $(0,0,0)$  2  $(0,0,0)$  2  $(2,2,0)$  2  $(2,2,0)$  2 &  1.622   0.984   \\ 
  $(0,0,0)$  2  $(0,0,0)$  2  $(2,2,0)$  3  $(2,2,0)$  3 &  1.621   0.981   \\ 
  $(0,0,0)$  3  $(0,0,0)$  3  $(2,2,0)$  1  $(2,2,0)$  1 &  1.544   0.967   \\ 
  $(0,0,0)$  3  $(0,0,0)$  3  $(2,2,0)$  2  $(2,2,0)$  2 &  1.621   0.981   \\ 
  $(0,0,0)$  3  $(0,0,0)$  3  $(2,2,0)$  3  $(2,2,0)$  3 &  1.622   0.984   \\ 
  $(0,0,0)$  1  $(0,0,0)$  1  $(3,0,0)$  1  $(3,0,0)$  1 &  1.753   1.007   \\ 
  $(0,0,0)$  1  $(0,0,0)$  1  $(3,0,0)$  2  $(3,0,0)$  2 &  1.828   1.021   \\ 
  $(0,0,0)$  1  $(0,0,0)$  1  $(3,0,0)$  3  $(3,0,0)$  3 & 1.719   0.999   \\ 
  $(0,0,0)$  2  $(0,0,0)$  2  $(3,0,0)$  1  $(3,0,0)$  1 &  1.828   1.021   \\ 
  $(0,0,0)$  2  $(0,0,0)$  2  $(3,0,0)$  2  $(3,0,0)$  2 &  1.752   1.007   \\ 
  $(0,0,0)$  2  $(0,0,0)$  2  $(3,0,0)$  3  $(3,0,0)$  3 &  1.787   1.013   \\ 
  $(0,0,0)$  3  $(0,0,0)$  3  $(3,0,0)$  1  $(3,0,0)$  1 &  1.719   0.999   \\ 
  $(0,0,0)$  3  $(0,0,0)$  3  $(3,0,0)$  2  $(3,0,0)$  2 &  1.787   1.013   \\ 
  $(0,0,0)$  3  $(0,0,0)$  3  $(3,0,0)$  3  $(3,0,0)$  3 &  1.735   1.003   \\ 
  $(0,0,0)$  1  $(0,0,0)$  1  $(3,1,0)$  1  $(3,1,0)$  1 &  1.671   0.991   \\ 
  $(0,0,0)$  1  $(0,0,0)$  1  $(3,1,0)$  2  $(3,1,0)$  2 &  1.724   1.001   \\ 
  $(0,0,0)$  1  $(0,0,0)$  1  $(3,1,0)$  3  $(3,1,0)$  3 &  1.719   0.999   \\ 
  $(0,0,0)$  2  $(0,0,0)$  2  $(3,1,0)$  1  $(3,1,0)$  1 &  1.724   1.001   \\ 
  $(0,0,0)$  2  $(0,0,0)$  2  $(3,1,0)$  2  $(3,1,0)$  2 &  1.667   0.991   \\ 
  $(0,0,0)$  2  $(0,0,0)$  2  $(3,1,0)$  3  $(3,1,0)$  3 &  1.787   1.012   \\ 
  $(0,0,0)$  3  $(0,0,0)$  3  $(3,1,0)$  1  $(3,1,0)$  1 &  1.585   0.975   \\ 
  $(0,0,0)$  3  $(0,0,0)$  3  $(3,1,0)$  2  $(3,1,0)$  2 &  1.621   0.981   \\ 
  $(0,0,0)$  3  $(0,0,0)$  3  $(3,1,0)$  3  $(3,1,0)$  3 &  1.655   0.989   \\ 
\hline     
\hline 
\end{tabular} 
\end{minipage}
\end{tabular} 
\end{table} 

\clearpage

\begin{table} 
\begin{tabular}{cc} 
\begin{minipage}{1.0\hsize}
\begin{center}
\label{WrLa3all_offdiagonal2} 
\caption{
Effective interactions in the cGW-SIC for three-band hamiltonian of La$_2$CuO$_4$ (in eV).
Notations are the same as Table S8.
}
\ 
\begin{tabular}{c|c} 
\hline \hline \\ [-8pt]
  $(0,0,0)$  1  $(0,0,0)$  1  $(3,2,0)$  1  $(3,2,0)$  1 &  1.521   0.963   \\ 
  $(0,0,0)$  1  $(0,0,0)$  1  $(3,2,0)$  2  $(3,2,0)$  2 &  1.544   0.967   \\ 
  $(0,0,0)$  1  $(0,0,0)$  1  $(3,2,0)$  3  $(3,2,0)$  3 &  1.585   0.975   \\ 
  $(0,0,0)$  2  $(0,0,0)$  2  $(3,2,0)$  1  $(3,2,0)$  1 &  1.543   0.967   \\ 
  $(0,0,0)$  2  $(0,0,0)$  2  $(3,2,0)$  2  $(3,2,0)$  2 &  1.514   0.962   \\ 
  $(0,0,0)$  2  $(0,0,0)$  2  $(3,2,0)$  3  $(3,2,0)$  3 &  1.621   0.981   \\ 
  $(0,0,0)$  3  $(0,0,0)$  3  $(3,2,0)$  1  $(3,2,0)$  1 &  1.465   0.952   \\ 
  $(0,0,0)$  3  $(0,0,0)$  3  $(3,2,0)$  2  $(3,2,0)$  2 &  1.479   0.954   \\ 
  $(0,0,0)$  3  $(0,0,0)$  3  $(3,2,0)$  3  $(3,2,0)$  3 &  1.511   0.962   \\ 
  $(0,0,0)$  1  $(0,0,0)$  1  $(3,3,0)$  1  $(3,3,0)$  1 &  1.453   0.950   \\ 
  $(0,0,0)$  1  $(0,0,0)$  1  $(3,3,0)$  2  $(3,3,0)$  2 &  1.465   0.952   \\ 
  $(0,0,0)$  1  $(0,0,0)$  1  $(3,3,0)$  3  $(3,3,0)$  3 &  1.465   0.952   \\ 
  $(0,0,0)$  2  $(0,0,0)$  2  $(3,3,0)$  1  $(3,3,0)$  1 &  1.465   0.952   \\ 
  $(0,0,0)$  2  $(0,0,0)$  2  $(3,3,0)$  2  $(3,3,0)$  2 &  1.445   0.949   \\ 
  $(0,0,0)$  2  $(0,0,0)$  2  $(3,3,0)$  3  $(3,3,0)$  3 &  1.479   0.954   \\ 
  $(0,0,0)$  3  $(0,0,0)$  3  $(3,3,0)$  1  $(3,3,0)$  1 &  1.465   0.952   \\ 
  $(0,0,0)$  3  $(0,0,0)$  3  $(3,3,0)$  2  $(3,3,0)$  2 &  1.479   0.954   \\ 
  $(0,0,0)$  3  $(0,0,0)$  3  $(3,3,0)$  3  $(3,3,0)$  3 &  1.445   0.949   \\ 
   $(0,0,0)$  1  $(0,0,0)$  1  $(0.5,0.5,1)$  1  $(0.5,0.5,1)$  1   & 2.076   1.112   \\ 
  $(0,0,0)$  1  $(0,0,0)$  1  $(0.5,0.5,1)$  2  $(0.5,0.5,1)$  2   & 2.128   1.130   \\ 
  $(0,0,0)$  1  $(0,0,0)$  1  $(0.5,0.5,1)$  3  $(0.5,0.5,1)$  3   & 2.128   1.130   \\ 
  $(0,0,0)$  2  $(0,0,0)$  2  $(0.5,0.5,1)$  1  $(0.5,0.5,1)$  1   & 1.917   1.068   \\ 
  $(0,0,0)$  2  $(0,0,0)$  2  $(0.5,0.5,1)$  2  $(0.5,0.5,1)$  2   & 2.063   1.111   \\ 
  $(0,0,0)$  2  $(0,0,0)$  2  $(0.5,0.5,1)$  3  $(0.5,0.5,1)$  3   & 1.952   1.078   \\ 
  $(0,0,0)$  3  $(0,0,0)$  3  $(0.5,0.5,1)$  1  $(0.5,0.5,1)$  1   & 1.917   1.068   \\ 
  $(0,0,0)$  3  $(0,0,0)$  3  $(0.5,0.5,1)$  2  $(0.5,0.5,1)$  2   & 1.952   1.078   \\ 
  $(0,0,0)$  3  $(0,0,0)$  3  $(0.5,0.5,1)$  3  $(0.5,0.5,1)$  3   & 2.063   1.111   \\ 
  $(0,0,0)$  1  $(0,0,0)$  1  $(1.5,0.5,1)$  1  $(1.5,0.5,1)$  1   & 1.753   1.026   \\ 
  $(0,0,0)$  1  $(0,0,0)$  1  $(1.5,0.5,1)$  2  $(1.5,0.5,1)$  2   & 1.917   1.068   \\ 
  $(0,0,0)$  1  $(0,0,0)$  1  $(1.5,0.5,1)$  3  $(1.5,0.5,1)$  3   & 1.771   1.031   \\ 
  $(0,0,0)$  2  $(0,0,0)$  2  $(1.5,0.5,1)$  1  $(1.5,0.5,1)$  1   & 1.607   0.992   \\ 
  $(0,0,0)$  2  $(0,0,0)$  2  $(1.5,0.5,1)$  2  $(1.5,0.5,1)$  2   & 1.746   1.026   \\ 
  $(0,0,0)$  2  $(0,0,0)$  2  $(1.5,0.5,1)$  3  $(1.5,0.5,1)$  3   & 1.615   0.993   \\ 
  $(0,0,0)$  3  $(0,0,0)$  3  $(1.5,0.5,1)$  1  $(1.5,0.5,1)$  1   & 1.669   1.008   \\ 
  $(0,0,0)$  3  $(0,0,0)$  3  $(1.5,0.5,1)$  2  $(1.5,0.5,1)$  2   & 1.796   1.038   \\ 
  $(0,0,0)$  3  $(0,0,0)$  3  $(1.5,0.5,1)$  3  $(1.5,0.5,1)$  3   & 1.739   1.026   \\ 
  $(0,0,0)$  1  $(0,0,0)$  1  $(1.5,1.5,1)$  1  $(1.5,1.5,1)$  1   & 1.583   0.988   \\ 
  $(0,0,0)$  1  $(0,0,0)$  1  $(1.5,1.5,1)$  2  $(1.5,1.5,1)$  2   & 1.669   1.008   \\ 
  $(0,0,0)$  1  $(0,0,0)$  1  $(1.5,1.5,1)$  3  $(1.5,1.5,1)$  3   & 1.669   1.008   \\ 
  $(0,0,0)$  2  $(0,0,0)$  2  $(1.5,1.5,1)$  1  $(1.5,1.5,1)$  1   & 1.493   0.968   \\ 
\hline     
\hline 
\end{tabular} 
\end{center}
\end{minipage}
\end{tabular} 
\end{table} 

\clearpage

\begin{table*}[h] 
\label{WrLa3all_offdiagonal3} 
\caption{
Effective interactions in the cGW-SIC for three-band hamiltonian of La$_2$CuO$_4$ (in eV).
Notations are the same as Table S8.
}
\ 
\begin{tabular}{c|c} 
\hline \hline \\ [-8pt]
  $(0,0,0)$  2  $(0,0,0)$  2  $(1.5,1.5,1)$  2  $(1.5,1.5,1)$  2   & 1.574   0.988   \\ 
  $(0,0,0)$  2  $(0,0,0)$  2  $(1.5,1.5,1)$  3  $(1.5,1.5,1)$  3   & 1.550   0.980   \\ 
  $(0,0,0)$  3  $(0,0,0)$  3  $(1.5,1.5,1)$  1  $(1.5,1.5,1)$  1   & 1.493   0.968   \\ 
  $(0,0,0)$  3  $(0,0,0)$  3  $(1.5,1.5,1)$  2  $(1.5,1.5,1)$  2   & 1.550   0.980   \\ 
  $(0,0,0)$  3  $(0,0,0)$  3  $(1.5,1.5,1)$  3  $(1.5,1.5,1)$  3   & 1.574   0.988   \\ 
  $(0,0,0)$  1  $(0,0,0)$  1  $(2.5,0.5,1)$  1  $(2.5,0.5,1)$  1   & 1.523   0.973   \\ 
  $(0,0,0)$  1  $(0,0,0)$  1  $(2.5,0.5,1)$  2  $(2.5,0.5,1)$  2   & 1.607   0.992   \\ 
  $(0,0,0)$  1  $(0,0,0)$  1  $(2.5,0.5,1)$  3  $(2.5,0.5,1)$  3   & 1.526   0.973   \\ 
  $(0,0,0)$  2  $(0,0,0)$  2  $(2.5,0.5,1)$  1  $(2.5,0.5,1)$  1   & 1.487   0.964   \\ 
  $(0,0,0)$  2  $(0,0,0)$  2  $(2.5,0.5,1)$  2  $(2.5,0.5,1)$  2   & 1.518   0.972   \\ 
  $(0,0,0)$  2  $(0,0,0)$  2  $(2.5,0.5,1)$  3  $(2.5,0.5,1)$  3   & 1.488   0.964   \\ 
  $(0,0,0)$  3  $(0,0,0)$  3  $(2.5,0.5,1)$  1  $(2.5,0.5,1)$  1   & 1.481   0.964   \\ 
  $(0,0,0)$  3  $(0,0,0)$  3  $(2.5,0.5,1)$  2  $(2.5,0.5,1)$  2   & 1.550   0.980   \\ 
  $(0,0,0)$  3  $(0,0,0)$  3  $(2.5,0.5,1)$  3  $(2.5,0.5,1)$  3   & 1.510   0.971   \\ 
  $(0,0,0)$  1  $(0,0,0)$  1  $(2.5,1.5,1)$  1  $(2.5,1.5,1)$  1   & 1.442   0.957   \\ 
  $(0,0,0)$  1  $(0,0,0)$  1  $(2.5,1.5,1)$  2  $(2.5,1.5,1)$  2   & 1.493   0.968   \\ 
  $(0,0,0)$  1  $(0,0,0)$  1  $(2.5,1.5,1)$  3  $(2.5,1.5,1)$  3   & 1.481   0.964   \\ 
  $(0,0,0)$  2  $(0,0,0)$  2  $(2.5,1.5,1)$  1  $(2.5,1.5,1)$  1   & 1.415   0.950   \\ 
  $(0,0,0)$  2  $(0,0,0)$  2  $(2.5,1.5,1)$  2  $(2.5,1.5,1)$  2   & 1.435   0.956   \\ 
  $(0,0,0)$  2  $(0,0,0)$  2  $(2.5,1.5,1)$  3  $(2.5,1.5,1)$  3   & 1.449   0.956   \\ 
  $(0,0,0)$  3  $(0,0,0)$  3  $(2.5,1.5,1)$  1  $(2.5,1.5,1)$  1   & 1.390   0.945   \\ 
  $(0,0,0)$  3  $(0,0,0)$  3  $(2.5,1.5,1)$  2  $(2.5,1.5,1)$  2   & 1.426   0.953   \\ 
  $(0,0,0)$  3  $(0,0,0)$  3  $(2.5,1.5,1)$  3  $(2.5,1.5,1)$  3   & 1.432   0.956   \\ 
  $(0,0,0)$  1  $(0,0,0)$  1  $(2.5,2.5,1)$  1  $(2.5,2.5,1)$  1   & 1.363   0.940   \\ 
  $(0,0,0)$  1  $(0,0,0)$  1  $(2.5,2.5,1)$  2  $(2.5,2.5,1)$  2   & 1.390   0.945   \\ 
  $(0,0,0)$  1  $(0,0,0)$  1  $(2.5,2.5,1)$  3  $(2.5,2.5,1)$  3   & 1.390   0.945   \\ 
  $(0,0,0)$  2  $(0,0,0)$  2  $(2.5,2.5,1)$  1  $(2.5,2.5,1)$  1   & 1.343   0.935   \\ 
  $(0,0,0)$  2  $(0,0,0)$  2  $(2.5,2.5,1)$  2  $(2.5,2.5,1)$  2   & 1.354   0.939   \\ 
  $(0,0,0)$  2  $(0,0,0)$  2  $(2.5,2.5,1)$  3  $(2.5,2.5,1)$  3   & 1.367   0.940   \\ 
  $(0,0,0)$  3  $(0,0,0)$  3  $(2.5,2.5,1)$  1  $(2.5,2.5,1)$  1   & 1.343   0.935   \\ 
  $(0,0,0)$  3  $(0,0,0)$  3  $(2.5,2.5,1)$  2  $(2.5,2.5,1)$  2   & 1.367   0.940   \\ 
  $(0,0,0)$  3  $(0,0,0)$  3  $(2.5,2.5,1)$  3  $(2.5,2.5,1)$  3   & 1.354   0.939   \\ 
 \hline \\ [-8pt]   
   $(0,1,0)$  3  $(0,0,0)$  1  $(0,1,0)$  3  $(0,1,0)$  3  & 0.175   0.103   \\ 
  $(0,-1,0)$  1  $(0,0,0)$  3  $(0,0,0)$  3  $(0,0,0)$  3 &  0.175   0.103   \\ 
  $(1,-1,0)$  2  $(0,0,0)$  3  $(0,-1,0)$  1  $(0,-1,0)$  1 &  0.286   0.099   \\ 
 $(-1,1,0)$  3  $(0,0,0)$  2 $(-1,0,0)$  1 $(-1,0,0)$  1 &  0.286   0.099   \\ 
  $(0,0,0)$  3  $(0,0,0)$  3  $(0,-1,0)$  1  $(0,0,0)$  3 &  0.175   0.103   \\ 
  $(0,0,0)$  3  $(0,0,0)$  3  $(0,0,0)$  3  $(0,-1,0)$  1 &  0.175   0.103   \\ 
  $(0,0,0)$  1  $(0,0,0)$  1  $(0,1,0)$  3  $(1,0,0)$  2  & 0.286   0.099   \\ 
  $(0,0,0)$  1  $(0,0,0)$  1  $(1,0,0)$  2  $(0,1,0)$  3 &  0.286   0.099   \\ 
\hline     
\hline 
\end{tabular} 
\end{table*}
